\documentclass[10pt,journal,compsoc]{IEEEtran}

\ifCLASSOPTIONcompsoc
  \usepackage[nocompress,noadjust]{cite} %
\else
  \usepackage{cite}
\fi

\ifCLASSINFOpdf
  \usepackage[pdftex]{graphicx}
  \graphicspath{{figures/}{graphs/}}
  \DeclareGraphicsExtensions{.pdf,.jpeg,.png}
\else
\fi
\newlength{\FigWidth}
\ifCLASSOPTIONonecolumn
\else
\fi

\usepackage{todonotes}
\usepackage{xspace}

\usepackage[pscoord]{eso-pic}%
\newcommand{\placetextbox}[3]{%
  \setbox0=\hbox{#3}%
  \AddToShipoutPictureFG*{%
    \put(\LenToUnit{#1\paperwidth},\LenToUnit{#2\paperheight}){\vtop{{\null}\makebox[0pt][c]{#3}}}%
  }%
}%

\usepackage{booktabs}
\ifCLASSOPTIONonecolumn
\else
   
\fi

\newcommand{\etal}{et al.\@\xspace} %
\newcommand{\eg}{e.g.\@\xspace} %

\newcommand{\ParLabel}[1]{\vspace{\topsep}\noindent{\sf\bfseries\small #1}}
\newcommand{\ParLabelFirst}[1]{\noindent{\sf\bfseries\small #1}}

\newcommand{\ParDesFirst}[1]{{\em #1}:}
\newcommand{\ParDes}[1]{\vspace{\topsep}\noindent{\em #1}:}

\begin{document}

\title{The Effect of Focal Distance, Age, and Brightness on Near-Field Augmented Reality Depth Matching}

\author{Gujot~Singh,~\IEEEmembership{Member,~IEEE,}
        Stephen~R.~Ellis,
        and~J.~Edward~Swan~II,~\IEEEmembership{Senior Member,~IEEE}%
\IEEEcompsocitemizethanks{%
\IEEEcompsocthanksitem Gurjot Singh is with Fairleigh Dickinson University. \protect\\
E-mail: gurjot@acm.org.
\IEEEcompsocthanksitem Stephen R. Ellis a Consultant, and was formally with NASA Ames Research Center.  E-mail: ellisstephenr3@gmail.com.
\IEEEcompsocthanksitem J. Edward Swan~II is with Mississippi State University. \protect\\
E-mail: swan@acm.org.}%
\thanks{Manuscript received XXXX; revised XXXX.}}

\markboth{IEEE Electronic Preprint (Under Review)}%
{Singh \MakeLowercase{\textit{et al.}}: Effect of Focal Distance, Age, and Brightness}

\IEEEtitleabstractindextext{%
\begin{abstract}
Many augmented reality (AR) applications operate within near-field reaching distances, and require matching the depth of a virtual object with a real object.  The accuracy of this matching was measured in three experiments, which examined the effect of focal distance, age, and brightness, within distances of 33.3 to 50~cm, using a custom-built AR haploscope.  Experiment~I examined the effect of focal demand, at the levels of collimated (infinite focal distance), consistent with other depth cues, and at the midpoint of reaching distance.  Observers were too young to exhibit age-related reductions in accommodative ability.  The depth matches of collimated targets were increasingly overestimated with increasing distance, consistent targets were slightly underestimated, and midpoint targets were accurately estimated.  Experiment~II replicated Experiment~I, with older observers.  Results were similar to Experiment~I.  Experiment~III replicated Experiment~I with dimmer targets, using young observers.  Results were again consistent with Experiment~I, except that both consistent and midpoint targets were accurately estimated.  In all cases, collimated results were explained by a model, where the collimation biases the eyes' vergence angle outwards by a constant amount.  Focal demand and brightness affect near-field AR depth matching, while age-related reductions in accommodative ability have no effect. 
\end{abstract}

}

\maketitle
\IEEEdisplaynontitleabstractindextext
\IEEEpeerreviewmaketitle

\placetextbox{0.285}{0.055}{\parbox{\columnwidth}{\tiny\copyright 2017 IEEE. Personal use of this material is permitted. Permission from IEEE must be obtained for all other uses, in any current or future media, including reprinting/republishing this material for advertising or promotional purposes, creating new collective works, for resale or redistribution to servers or lists, or reuse of any copyrighted component of this work in other works.}}

\IEEEraisesectionheading{%
\section {Introduction}%
\label{s:intro}}

\IEEEPARstart{M}{any} compelling applications of augmented reality (AR) require interacting with real and virtual objects at reaching distances.  Some examples include image-guided medical procedures (\eg, Kersten-Oertel \etal\cite{kersten-oertel:2013}), manufacturing (\eg, Curtis \etal\cite{curtis:1998}), and maintenance (\eg, Henderson and Feiner \cite{henderson:2009}).  Among the factors that determine success is the accuracy with which observers can match the distance of a real object to an AR-presented virtual object.  For example, a surgeon may need to cut to the depth indicated by an AR-presented tumor, or place a needle within the tumor.  In order for AR to be useful for image-guided surgery of the brain, Edwards \etal\cite{edwards:2000} found that surgeons must be able to place a scalpel with a tolerance of 1 mm; and, in order for AR to be useful for a type of radiation therapy, Krempien \etal~\cite{krempien:2008} found that a needle must be placed with a tolerance of 1 mm.

\begin{figure}[!t]
\centering
\includegraphics[width=\FigWidth]{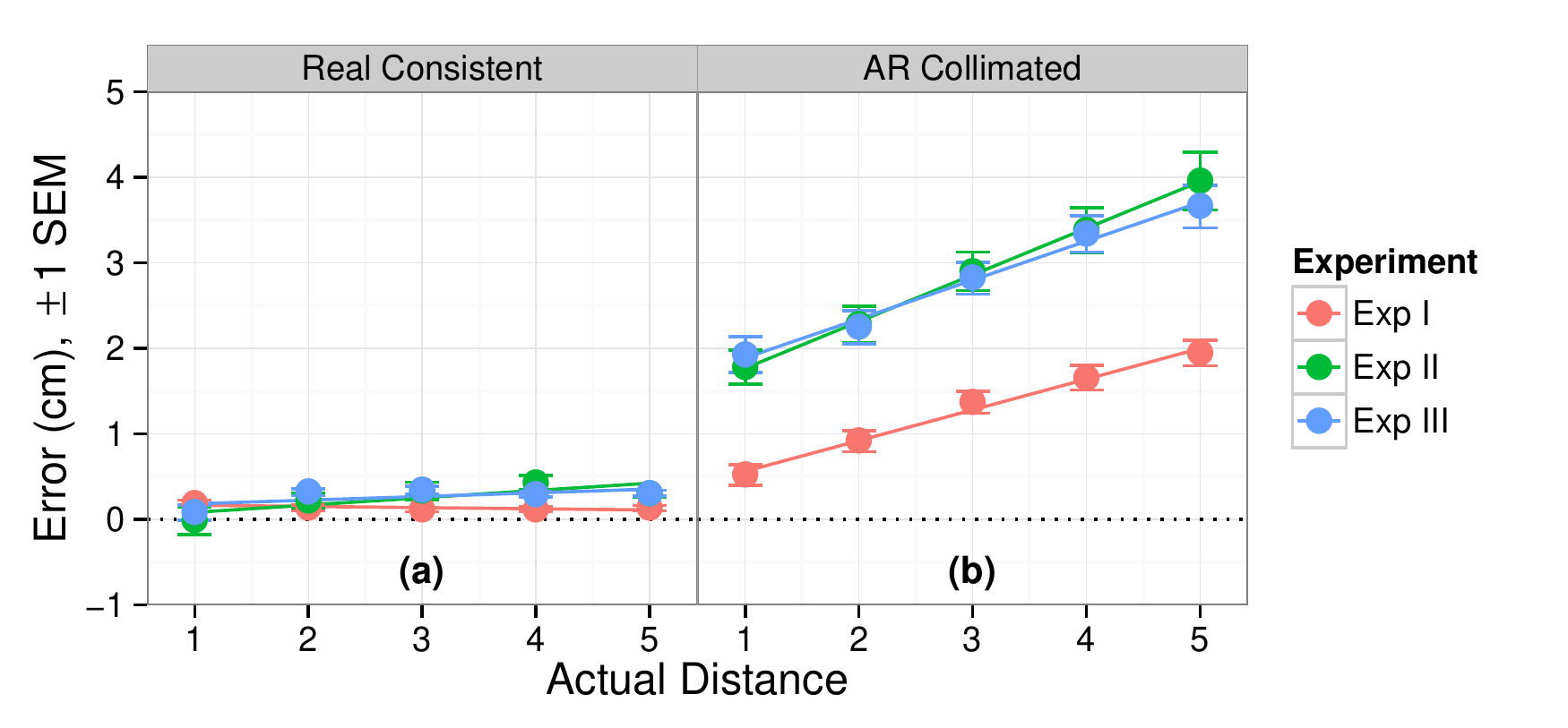}
\caption{The perceptual matching depth judgments from Swan~\etal~\cite{swan:2015}.  For Experiment I, the actual distances were 34, 38, 42, 46, and 50~cm, while for Experiments II and III the actual distances were 55, 63, 71, 79, and 87\% of each observer's maximum reach.}
\label{f:swan2015}
\end{figure}

In previous work motivated by this topic, Swan~\etal\cite{swan:2015} reported initial efforts to measure the accuracy of AR depth matching.  An optical see-through AR display was used, and reaching distances of 24 to 56~cm were examined.  The depth judgment was \emph{perceptual matching}, where observers adjusted a pointing object in depth, until they judged it to be the same distance from themselves as a target object.  Fig.~\ref{f:swan2015} summarizes these results, which were collected across three experiments.  The pointer was always a real object, and therefore its distance from the observer could be objectively measured in the real world.  In Fig.~\ref{f:swan2015}, the $x$-axis is the actual depth of the target object, and the $y$-axis is the depth error of the pointer.  Here, $\mbox{\it error} = 0$ indicates that observers placed the pointer at the same depth as the target object; $\mbox{\it error} > 0$ indicates \emph{overestimated} depth matches, where observers placed the pointer farther in depth than the target object; and $\mbox{\it error} < 0$ indicates \emph{underestimated} depth matches, where observers placed the pointer closer than the target object.  As a control condition, Swan~\etal\cite{swan:2015} examined the accuracy of matching a \emph{real} target object, and found accuracies of 1.4 to 2.7 mm (Fig.~\ref{f:swan2015}a, the \emph{real consistent} condition).  However, when they examined matching a \emph{virtual} target object, they found that observers systematically overestimated the matching distance, ranging from 0.5~cm at near distances to 4.0~cm at far distances (Fig.~\ref{f:swan2015}b, the \emph{AR collimated} condition).  Therefore, as illustrated in Fig.~\ref{f:swan2015}, there was a significant difference in depth matching real and virtual targets.  

\begin{figure}[!t]
\centering
\includegraphics[width=0.7\FigWidth]{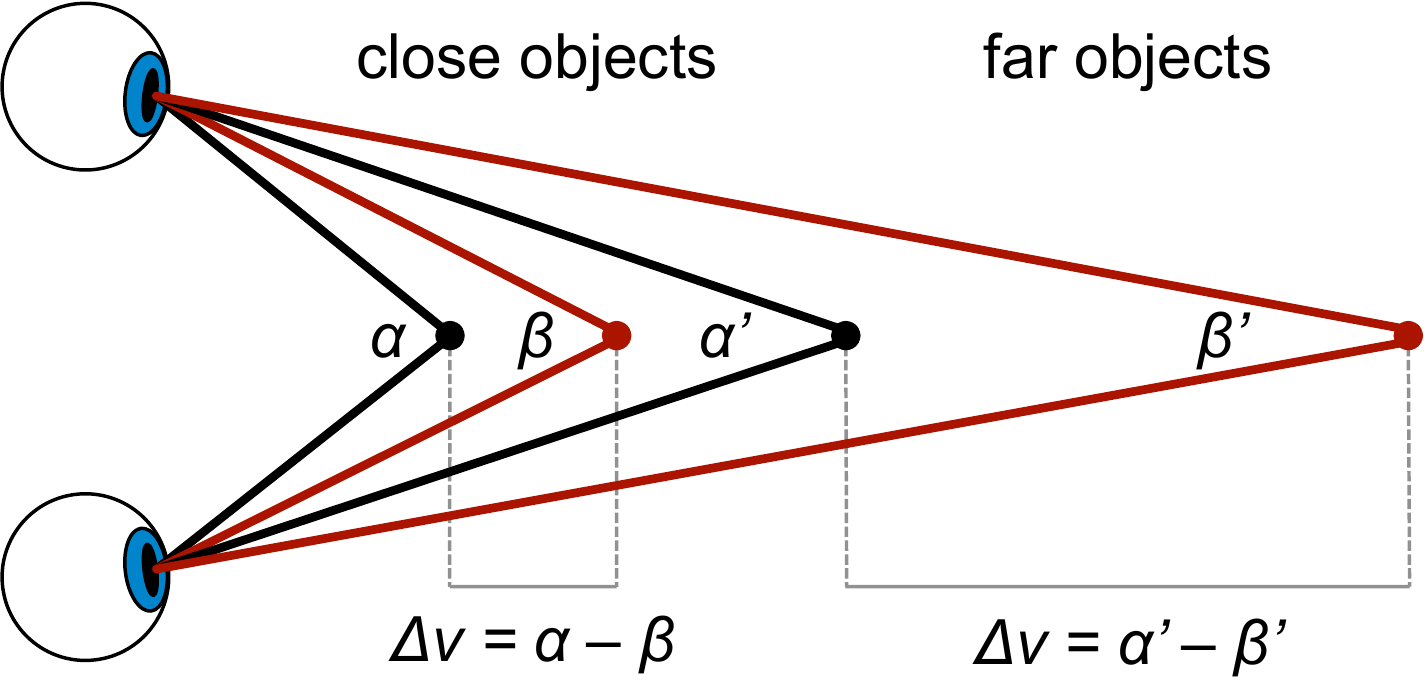}
\caption{The model that explains how a constant change in vergence angle, $\Delta v$, leads to matched distances of virtual objects (red: $\beta$, $\beta'$), relative to real objects (black: $\alpha$, $\alpha'$), that are increasingly overestimated with increasing distance.}
\label{f:model}
\end{figure}

Swan~\etal\cite{swan:2015} determined that the likely reason for these results was that their AR display used collimating optics, which present virtual objects focused at optical infinity.  They found the results to be very well described by a model where this collimation causes the eyes' vergence angle to rotate outwards by a constant amount.  Fig.~\ref{f:model} illustrates this model.  Let the black points labelled $\alpha$ and $\alpha'$ be two real objects, with the first located close to the observer, and the second located farther away.  And, let the red points labelled $\beta$ and $\beta'$ be two virtual objects, which are rendered to be the same distance as the real targets.  $\alpha$, $\alpha'$, $\beta$, and $\beta'$ also represent the angle of binocular parallax that the eyes make when the observer fixates on each object.  
Therefore, when fixating on the close real object, the angle of binocular parallax is $\alpha$, and if the fixation changes to the close virtual object, then the collimation causes the eyes to rotate outwards, reducing the angle to $\beta$. 
When placing the real pointer $\alpha$ at the same depth as the virtual target $\beta$, observers' eyes rotate inwards and outwards as they fixate between the two objects, and therefore observers perceive them to be located at the same depth.  
The model predicts that this change in vergence angle, $\Delta v = \alpha - \beta$, is constant for reaching distances.  Therefore, when fixating on the far target, $\alpha' < \alpha$, and this same change in vergence angle, $\Delta v = \alpha' - \beta'$, causes a larger depth distance between $\alpha'$ and $\beta'$ (Fig.~\ref{f:model}).  
This model explains three properties of Swan~\etal's~\cite{swan:2015} results (Fig.~\ref{f:swan2015}): (1) because the collimating optics cause the eyes to rotate outwards, the depth judgments of the virtual targets are overestimated relative to the real targets, (2) the amount of overestimation increases with increasing distance, and (3) the results are very well fit with a linear model.   

This analysis suggests that, for accurate depth placement, virtual objects need to be presented with a \emph{focal depth}---also termed \emph{accommodative demand}---that is consistent with their intended depth.  Then, the eyes' vergence angle will not be biased, and depth matches will be more accurate.  This paper reports three experiments that systematically examine this hypothesis.

However, it was not possible to conduct these experiments with the same AR display as Swan~\etal~\cite{swan:2015}.  That display, an NVIS Inc.\ nVisor ST60 model, contains unadjustable collimating optics, which always present virtual objects focused at optical infinity.  This is consistent with the vast majority of commercially available AR displays, almost all of which have a focal distance that is set at the factory, and unadjustable by the end user.\footnote{%
The authors of the current paper, who have been studying virtual and augmented reality for 8, 30, and 18 years, respectively, can only recall a single commercially-available AR display---the Microvision Nomad from the early 2000's--- which came with a focus adjustment knob (Gabbard \etal~\cite{gabbard:2017}).} %
Therefore, an \emph{augmented reality haploscope}---an AR display mounted on an optical workbench that allows accommodative demand and vergence angle to be independently and precisely adjusted---was developed and used for the experiments reported here.\footnote{%
Portions of these experiments are reported in a PhD dissertation (Singh~\cite{singh:2013}).
}%

\section {Background and Related Work}

\subsection{Depth Perception and Depth Cues}

The human visual system achieves a percept of perceived depth from \emph{depth cues}---sources of perceptual information related to depth.  At least nine depth cues have been identified (Cutting and Vishton \cite{cutting:1995}): 
\emph{occlusion} (a closer object occludes farther objects), 
\emph{binocular disparity} (an object projects to different locations on each retina), 
\emph{motion perspective} (objects at different distances from a moving observer have different apparent velocities), 
\emph{height in the visual field} (starting from the horizon, closer objects are lower in the visual field), 
\emph{relative size} (among objects of the same size, the farther object projects to a smaller retinal angle), 
\emph{accommodation} (the lens of the human eye changes shape to bring objects into focus), 
\emph{vergence} (the two eyes rotate to fixate on an object (Fig.~\ref{f:model})), 
\emph{relative density} (for a textured surface, at farther distances more objects are seen within a constant retinal angle), and
\emph{aerial perspective} (objects at great distances lose color saturation and contrast).

Depth cues differ in effectiveness based on various visual characteristics, such as scene content and distance from the observer.  Nagata \cite{nagata:1991}, and later Cutting and Vishton \cite{cutting:1995}, organized the relative effectiveness of different depth cues according to distance.  Within near-field reaching distances, they find that the operative depth cues, in approximate order of decreasing salience, are 
\emph{occlusion}, 
\emph{binocular disparity}, 
\emph{motion perspective}, 
\emph{relative size},  
\emph{accommodation and vergence}, and 
\emph{relative density}. 

Most of these depth cues can be categorized as \emph{retinal}, because the information from the cue comes from the visual scene sensed on the retina.  However, the cues of accommodation and vergence are \emph{extra-retinal}, because the cue information comes from sensors that detect the state of the muscles that control the lenses' shape and the eyes' vergence angle.  In principal, the extra-retinal cues cues could provide absolute egocentric depth information (Gillam \cite{gillam:1995}).  In contrast, retinal cues can only provide relative depth information between objects in the scene; these cues require an external reference to establish the scene's overall scale.  However, when combined with extra-retinal cues, and an observer's constant interpupillary distance, retinal cues can also provide absolute depth information (Bingham and Pagano \cite{bingham:1998}, Mon-Williams and Tresilian \cite{monwilliams:1999}).  In general, the way the human visual system combines information from different depth cues to produce a stable percept of distance is subtle and not fully understood, although many theories have been advanced and the collected evidence favors some theories over others (Landy \etal~\cite{landy:1995}, Singh~\cite{singh:2013}).

\subsection{Vergence and Accommodation}

Visual perception requires a rapid series of precise eye movements (Leigh and Zee \cite{leigh:2015}), including 
\emph{fixation} (hold an image steady on the fovea by minimizing eye movement),
\emph{saccadic} (quick movement that projects an object of interest to the fovea), 
\emph{smooth pursuit} (retain fixation on an object during smooth movement of either the object or head),
\emph{vestibular} (hold vision steady during head movements), and
\emph{vergence} (the two eyes rotate to fixate on an object of interest (Fig.~\ref{f:model})).  
When changing fixation from a far to a near object, the eyes converge, the lenses become thicker, and the pupils constrict.  These three actions---vergence, accommodation, and changing pupil size---are interlinked physiologically, and the mechanism of these three simultaneous reflexes is called the \emph{near triad}.  Because of the interlinkage, changes in either accommodation or vergence drive corresponding changes in the other two components of the triad (Semmlow and Hung \cite{semmlow:1983}).  Apart from the influence of accommodation and vergence, pupil diameter also changes according to scene illumination, becoming larger in dim settings and smaller in bright settings.  Although these illumination-driven changes in pupil diameter affect the eye's optical depth of field, and therefore could potentially affect accommodation, little effect of changing pupil diameter on accommodation has been observed (Ripps \etal \cite{ripps:1962}).  Therefore, in near field viewing, vergence and accommodation are the main depth reflexes, and the link between them is known as the \emph{vergence-accommodation reflex}.  Because of this reflex, accommodation and vergence operate in unison: changes in accommodation drive changes in vergence (\emph{accommodative vergence}), and changes in vergence drive changes in accommodation (\emph{vergence accommodation}) (Kersten and Legge \cite{kersten:1983}).  Therefore, the vergence reflex is driven both by binocular disparity (the eyes rotate to bring a fixated object to a level of zero binocular disparity), as well as accommodative vergence.  Likewise, the accommodation reflex is driven both by focal blur (the lenses adjust to minimize blur), as well as vergence accommodation (Mon-Williams and Tresilian \cite{monwilliams:2000}).   

\begin{figure}[!t]
\centering
\includegraphics[width=1\FigWidth]{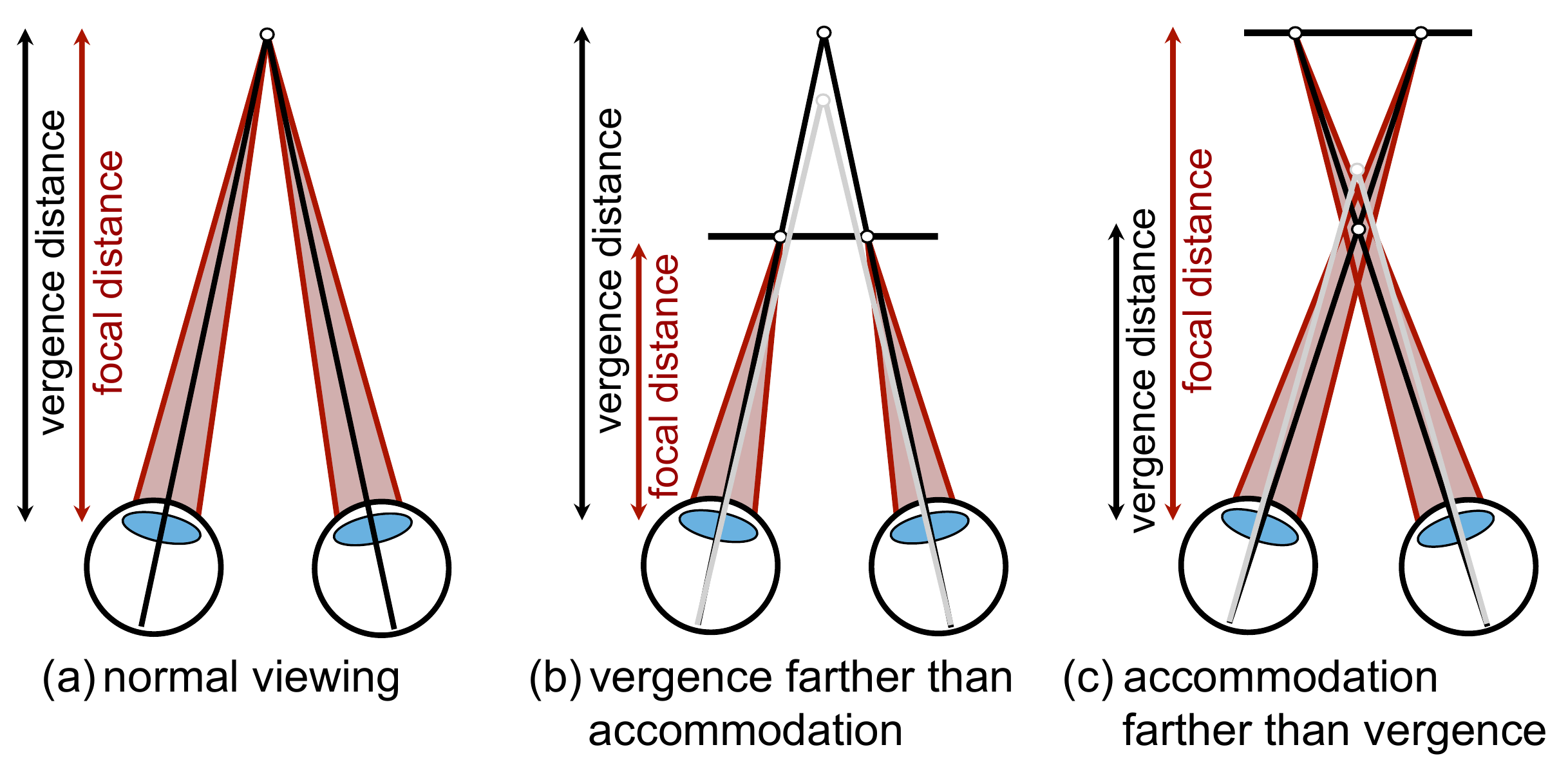}
\caption{The vergence-accommodation conflict, and its effect on perceived depth.  (a) In normal viewing of real world objects, the vergence distance, required for zero binocular disparity, is the same as the focal distance, required for minimal focal blur.  (b) When the vergence distance is farther than the focal distance, e.g. when viewing a virtual object beyond the surface of a stereo monitor, the vergence angle is biased inwards (grey lines), and the object is seen as closer than encoded by disparity.  (c) When the vergence distance is closer than the focal distance, e.g. when viewing a virtual object in front of the surface of a stereo monitor, the vergence angle is biased outwards (grey lines), and the object is seen as farther than encoded by disparity.} 
\label{f:verg-acc-con}
\end{figure}

Of course, the vergence-accommodation reflex is calibrated for viewing real world objects, which present consistent binocular disparity and focal blur cues (Fig.~\ref{f:verg-acc-con}a).  When viewing virtual objects, the binocular disparity and focal blur cues are often inconsistent, because the focal blur cue is fixed at the screen depth, while the depth of the binocular disparity cue varies, sometimes beyond the screen depth (Fig.~\ref{f:verg-acc-con}b), and sometimes in front (Fig.~\ref{f:verg-acc-con}c).  This is called the \emph{vergence-accommodation conflict}, and it is a ubiquitous aspect of all stereo displays with a single focal plane 
(Kruijff \etal \cite{kruijff:2010}).  The conflict causes visual fatigue (Gabbard \etal \cite{gabbard:2017}, Lambooij \etal \cite{lambooij:2009}), hinders visual performance (Hoffman \etal~\cite{hoffman:2008}), and biases depth perception towards the screen depth (Fig.~\ref{f:verg-acc-con}, Swenson~\cite{swenson:1932}, Mon-Williams and Tresilian~\cite{monwilliams:2000}).

The contribution of vergence to perceived depth depends upon various properties of the scene.  At near-field distances, vergence has been conclusively found to provide egocentric depth information (Brenner and Van Damme~\cite{brenner:1998}, Owens and Liebowitz~\cite{owens:1980}, Tresilian \etal~\cite{tresilian:1999}, Viguier \etal~\cite{viguier:2001}; Foley~\cite{foley:1980} provides a comprehensive review).  Although vergence in isolation is not a very accurate depth cue, observers are very sensitive to changes in vergence, which generally allows accurately matching the depth of one object with another (Brenner and Van Damme~\cite{brenner:1998}).  Each individual has a different vergence resting point---their \emph{dark vergence}---which is the vergence angle that their eyes assume when the controlling muscles are completely relaxed.  In low light conditions, the egocentric depth specified by vergence is biased towards each individual's dark vergence distance (Owens and Liebowitz~\cite{owens:1980}).  As a depth cue, vergence is most effective within near-field distances of 2 meters (Viguier \etal~\cite{viguier:2001}), a distance range that encompasses $\sim$90\% of vergence eye movements (Tresilian \etal~\cite{tresilian:1999}).  As other retinal depth cues become available, the contribution of vergence to perceived depth is reduced, but still present (Foley~\cite{foley:1980}).

As discussed above, accommodation influences perceived depth through the vergence-ac\-commodation reflex.  Although some studies have found evidence that accommodation alone can serve as a depth cue for some observers, these experiments require careful experimental setups to eliminate other depth cues, and the consensus remains that accommodation influences perceived depth through its effect on vergence (Mon-Williams and Tresilian~\cite{monwilliams:2000}).  Similar to dark vergence, each individual has a \emph{dark focus}---the distance their eyes focus when the controlling muscles are in a relaxed state (Iavecchia \etal~\cite{iavecchia:1988}).  The dark focus biases the eye's focal response, resulting in a number of perceptual consequences, including perceived depth (Roscoe~\cite{roscoe:1985}).  Generally, the dark focus and dark vergence distances vary independently, and for most individuals are not equal (Owens and Leibowitz~\cite{owens:1980}).

\subsection{Accommodation and Age}
\label{s:age}

Accommodative ability, the distance range within which a viewed object can be brought into clear focus, decreases with increasing age (Duane~\cite{duane:1912}), a condition known as \emph{presbyopia}.  It is primarily caused by hardening of the crystalline lens, although other physiological changes in the lens, connective tissue, and controlling muscles also play a role (Kasthurirangan and Glasser~\cite{kasthurirangan:2006}).  As measured by Duane~\cite{duane:1912}, presbyopia begins by the age of 12, but through the early 30's the loss is minuscule---the closest distance of clear focus declines from $\sim$8 to $\sim$13~cm.  However, the decline then accelerates, and by the age of 50 often surpasses 50~cm.  At some point in the 40's, the closest distance of clear focus often surpasses standard reading distance, and reading glasses are required.  By their mid-50's, most people have lost the ability to adjust the distance of clear focus.  

It seems reasonable that this loss of accommodative ability would have perceptual consequences, and indeed, older people are worse than younger people at many perceptual tasks (Bian and Andersen~\cite{bian:2013}).  However, accommodative vergence does not diminish with age; even as the visual system looses the ability to adjust accommodation, the eyes still verge properly in response to accommodative stimuli (Heron \etal~\cite{heron:2001}).
Because vergence is the primary source of depth information from the vergence-accommodation reflex (Mon-Williams and Tresilian~\cite{monwilliams:2000}), this suggests that depth perception could be unaffected by presbyopia.  Indeed, Bian and Andersen~\cite{bian:2013} found that, when making judgments of medium-field egocentric distances, older people (average 73.4 years) were \emph{more} accurate than younger people (average 22.5 years).  This is one of a series of recent studies that have found that older observers preserve their abilities in tasks related to distance perception (Bian and Andersen~\cite{bian:2013}). 

\subsection{Accommodation and Scene Flatness}

Another effect of the vergence-accommodation conflict in stereo displays is that the accommodative distance changes the perceived \emph{flatness} of the scene (Andersen \etal~\cite{andersen:1998}, Nagata~\cite{nagata:1991}, Singh~\cite{singh:2013}).  Specifically, when medium- to far-field scenery is shown on a display, but accommodative distance is in the near field, depth distances between scene objects are compressed, and the scene is perceived as being a flat window, positioned some depth distance from the observer.  However, when the same scene is shown with collimation, these depth distances are no longer compressed, and the scene objects appear to extend in depth, with some closer to the observer and others farther.  This is a reason why many augmented and virtual reality displays, especially those used for flight simulation and other far-field applications, present collimated light (Watt \etal~\cite{watt:2005}).  Likewise, the NVIS nVisor ST60 used by Swan~\etal \cite{swan:2015}, which also presents collimated light, was originally marketed for military training and forward observer tasks, which primarily involve medium- to far-field distances.  

\subsection{Depth Perception and Brightness}
\label{s:bright}

Among objects of the same size and distance, the brighter appear closer than the dimmer.  This principal has long been known in art, and is discussed by Leonardo Da Vinci in his \emph{Notebooks} (McCurdy~\cite{mccurdy:1938}).  The principal has been thoroughly studied, at both near field (Ashley~\cite{ashley:1898}, Farn\`{e}~\cite{farne:1977}) and medium field (Coules~\cite{coules:1955}) distances, and in both monocular and binocular conditions (Coules~\cite{coules:1955}).  In addition to brightness, the contrast between an object and the background also effects perceived depth, so a dark object against a light background can appear closer than an object with less contrast (Farn\`{e}~\cite{farne:1977}).  Among the theories that explain this effect are that brighter objects stimulate a larger area on the retina, and that brighter objects effect pupil size, which then biases other near triad reflexes.  

\subsection{Related Work in Augmented Reality}

To date, including Swan \etal~\cite{swan:2015}, only a small number of papers have examined near-field AR depth matching.
Ellis and Menges~\cite{ellis:1998} measured the effects of convergence, accommodation, observer age, viewing condition (monocular, biocular stereo, binocular stereo), and the presence of an occluding surface.  They found that accuracy is degraded by monocular viewing and an occluding surface. 
Using the same experimental setup, McCandless \etal~\cite{mccandless:2000} additionally studied motion parallax and latency in monocular viewing; they found reduced accuracy with increasing distance and latency.  
Singh \etal~\cite{singh:2010} found that an occluding surface has complex accuracy effects, 
and Rosa \etal~\cite{rosa:2016} found increased accuracy with redundant tactile feedback. 

\section {The Augmented Reality Haploscope}
\label{s:haplo}

As motivated in Section~\ref{s:intro}, an \emph{augmented reality haploscope} was designed and engineered.\footnote{%
Additional technical details, and a history of preliminary versions and design tradeoffs, can be found in Singh~\cite{singh:2013}.}
The design was loosely based on the AR haploscopes described by Rolland \etal\ \cite{rolland:1995} and Ellis and Menges~\cite{ellis:1998}, but similar designs have a long history in the study of depth perception (e.g., Swenson~\cite{swenson:1932}).  

\begin{figure}[!t]
\centering
\includegraphics[width=\FigWidth]{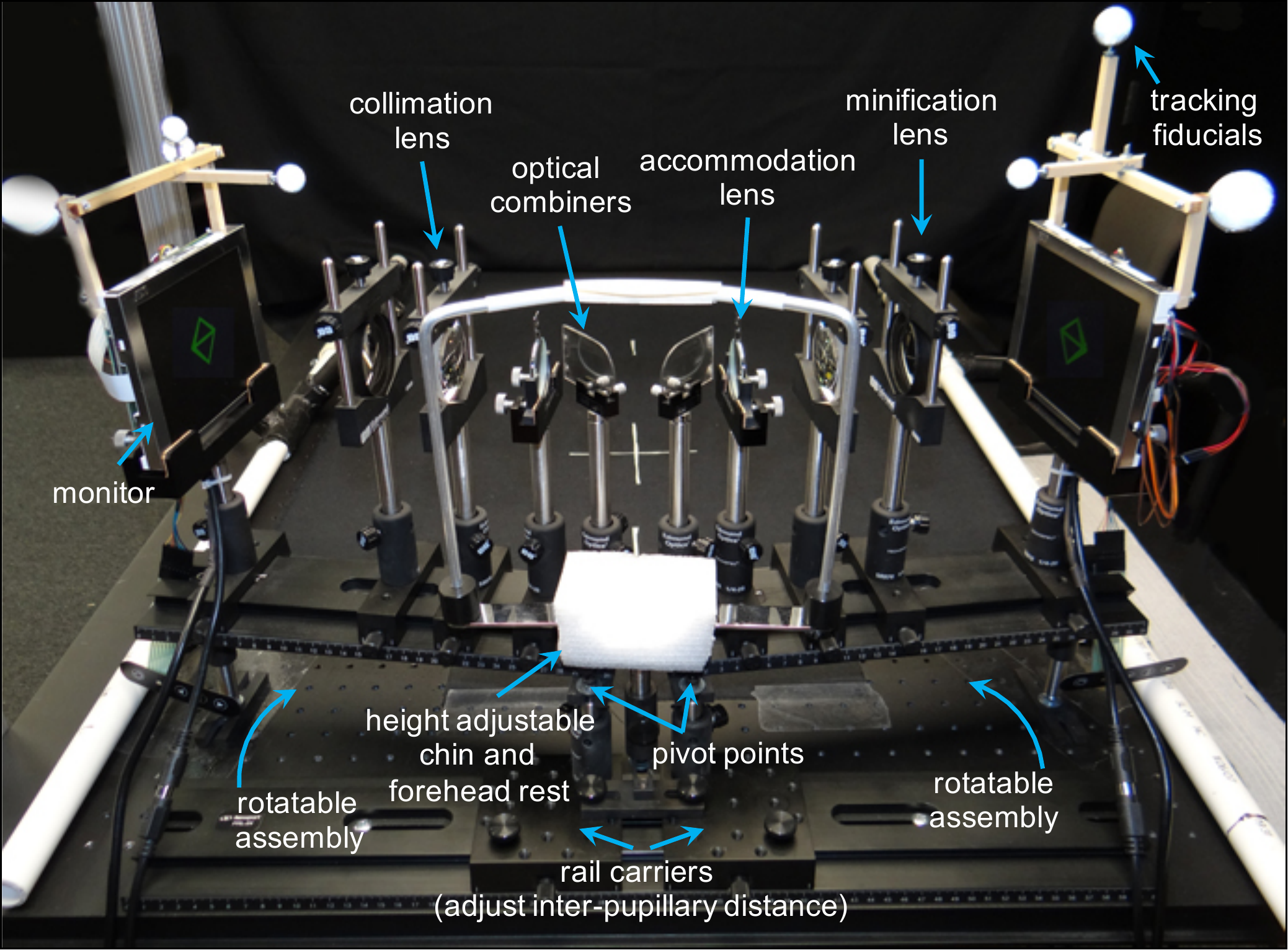}
\caption{The Augmented Reality (AR) Haploscope.  The physical design allows independent adjustment of vergence angle and focal distance.}
\label{f:haplo}
\end{figure}

Fig.~\ref{f:haplo} shows the AR haploscope.  The physical design has the following requirements: (1) provide a range of vergence angles and accommodative demands, (2) adjust to match a wide range of inter-pupillary distances, and (3) be rigid enough to resist inevitable bumps.  To achieve these, the device is mounted on an optical breadboard. 
The primary structure is built on three 
optical rails: two 12-inch rails serve as mounting bases for left-eye and right-eye optical systems, and both 12-inch rails are mounted on a 24-inch rail using 3-inch rail carriers, which can be adjusted to match the required inter-pupillary distance. 

\begin{figure}[!t]
\centering
\includegraphics[width=\FigWidth]{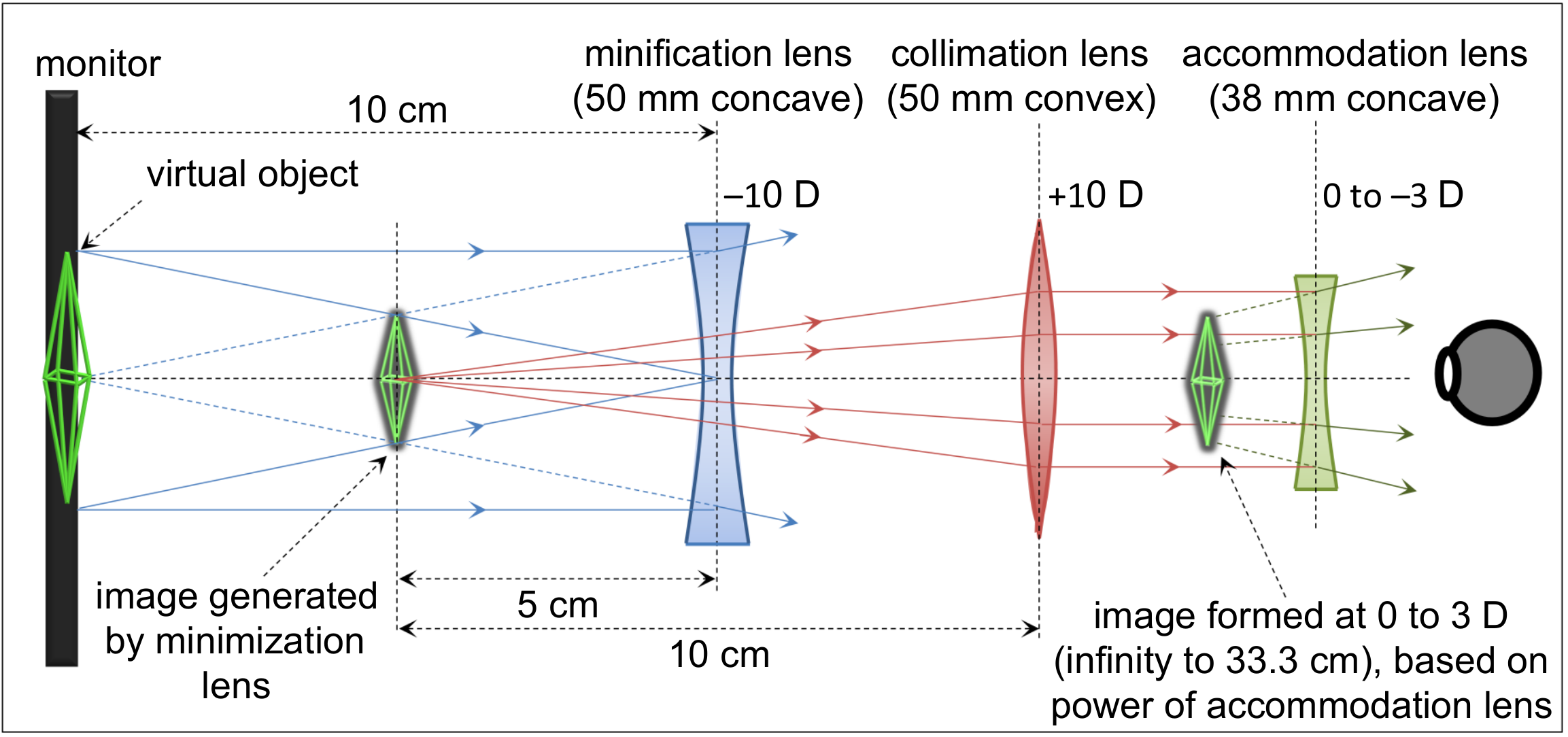}
\caption{The optical system of the AR haploscope.  Changing the accommodation lens changes the focal distance.}
\label{f:ray}
\end{figure}

The goal of each optical system is to collimate the generated image, so the image is located at optical infinity, or 0 diopters (D).  Then, the collimated image can either be left at optical infinity, or a negative power lens can reduce the focal distance.  Fig.~\ref{f:ray} shows the optical system.  The image is first generated by a monitor. 
Then, the image is minified by a $-$10~D concave lens;
without minification, only a small part of the monitor can be seen through the optical system.  As shown in Fig.~\ref{f:ray}, when this $-$10~D lens is placed 10~cm from the monitor, it creates a minified image at $-$5~cm.  This minified image is then collimated by a $+$10~D convex lens, 
positioned 10~cm from the image.  The collimated image is then passed through an \emph{accommodation lens}.  This comes from a standard optometric trial set; either a 0~D plain glass lens, which retains the collimation, or a negative power concave lens, which reduces the focal distance.  In the experiments reported here, the strongest accommodation lens used was $-$3~D, which resulted in a 33.3~cm focal distance.  
After generation, the images are reflected into the observers' eyes by 15\% reflective optical combiners, mounted at 45$^{\circ}$ directly in front of each eye.  
Fig.~\ref{f:haplo} shows the monitors; the minification, collimation, and accommodation lenses; and the optical combiners.

\begin{figure}[!t]
\centering
\includegraphics[width=0.9\FigWidth]{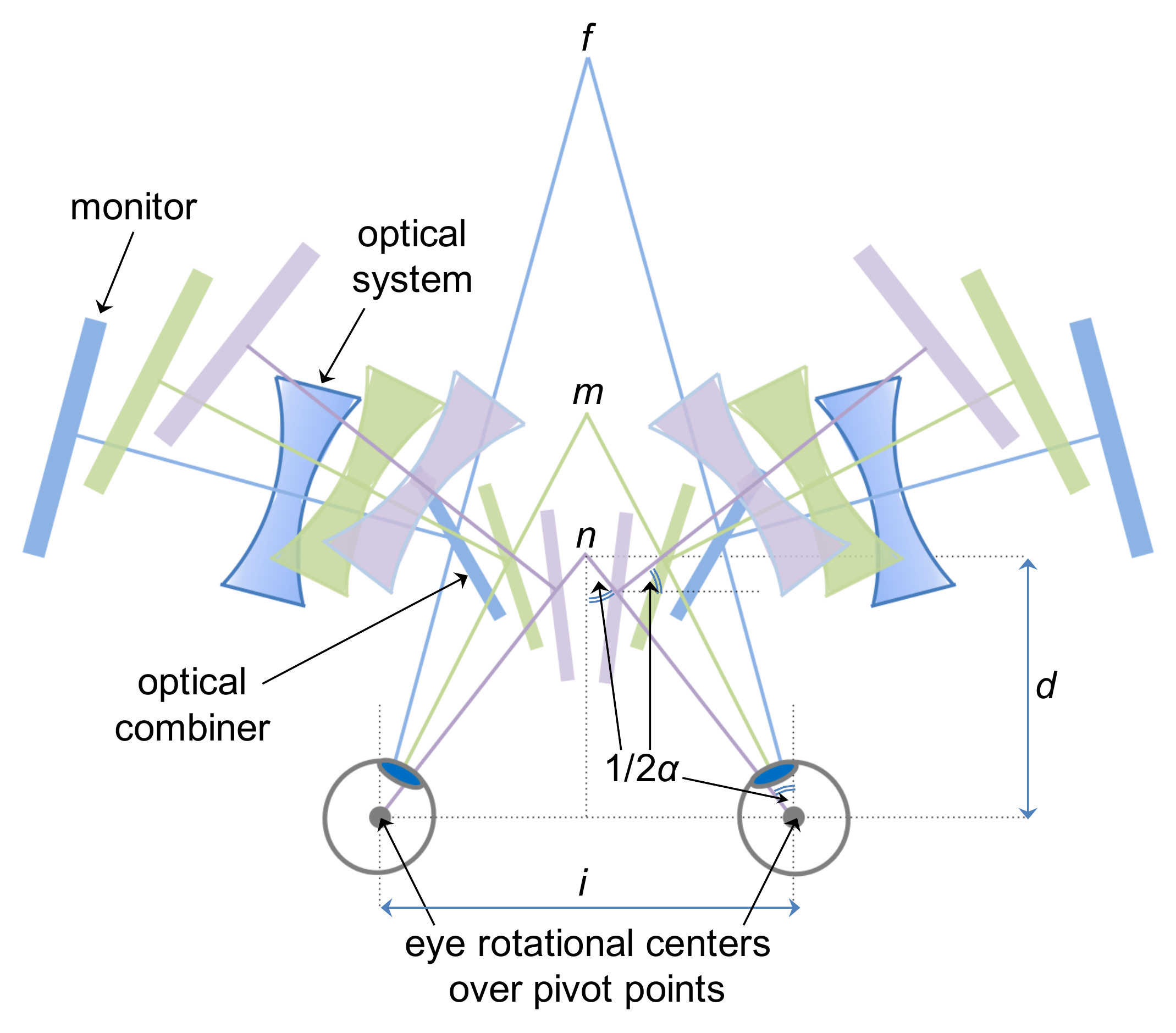}
\caption{Rotating the optical systems to match the correct vergence angle.}
\label{f:rotate}
\end{figure}

Fig.~\ref{f:rotate} illustrates how the haploscope matches different vergence angles.  The rail carriers are adjusted so that the distance between the pivot points matches the observer's inter-pupillary distance (Fig.~\ref{f:haplo}).  The chin and forehead rest is adjusted so that these pivot points are directly below the rotational centers of the observer's eyes.  As illustrated in Fig.~\ref{f:rotate}, when the left and right optical systems then rotate about the pivot points, for all convergence distances the view rays from the center of the two eyes stay in line with the principal axes of the optical systems.
This allows presenting a virtual object at any distance, near ($n$), medium ($m$), or far ($f$), while the observer's view rays continue to pass through the middle of the optical system, where optical distortion is minimized.  To display a target object at a specific distance, the optical systems are rotated to the matching convergence angle $1/2\alpha$ (Figs.~\ref{f:model}, \ref{f:rotate}); $1/2\alpha = \arctan(i/2d)$, where $i$ is the observer's interpupillary distance, and $d$ is the target distance.  The angle of each optical system is measured by a constellation of tracking fiducials attached to each monitor (Fig.~\ref{f:haplo}), which allows an ART TrackPack to measure the vergence angle to an accuracy of 0.01$^{\circ}$.

\section {Experiment~I: Accommodation}
\label{s:exI}

As discussed in Section~\ref{s:intro}, Swan \etal~\cite{swan:2015} hypothesized that the linearly increasing overestimation they found with collimated AR graphics (Fig.~\ref{f:swan2015}), was caused by the collimation biasing the eyes' vergence angle to rotate outwards by a constant amount (Fig.~\ref{f:model}).  The purposes of Experiment~I were to test aspects of this hypothesis, using the same matching task and within a similar range of near-field distances. 
Because Experiment~I used a different display---the AR haploscope---the first purpose (1) was to replicate the \emph{real consistent} and \emph{AR collimated} conditions of Swan \etal~\cite{swan:2015}.  If Experiment~I found similar results, that would suggest that these results generalize to AR more broadly, and are not specific to the NVIS display used by Swan \etal~\cite{swan:2015}.
The next purpose (2), the \emph{AR consistent} condition, was to test whether presenting AR objects at a focal distance that was \emph{consistent} with the distance specified by other depth cues, especially binocular disparity, would result in more accurate depth matches than what was seen in the AR collimated condition.  If the depth matches are more accurate, that would further support the hypothesis that collimated graphics bias the eyes' vergence angle outwards.
However, for many AR applications, always presenting virtual objects at consistent focal distances is unlikely to be practical.  Therefore, the final purpose (3), the \emph{AR midpoint} condition, was to test whether presenting AR objects at a focal distance equal to the \emph{midpoint} of the tested range would result in performance similar to the consistent condition.  If the performance is similar, this would suggest that, for accurate depth matching within reaching distances, the expense of making the focal demand consistent for every virtual object is not necessary.

\subsection {Method}

\subsubsection {Apparatus and Task}
\label{s:app}

\begin{figure}[!t]
\centering
\includegraphics[width=0.8\FigWidth]{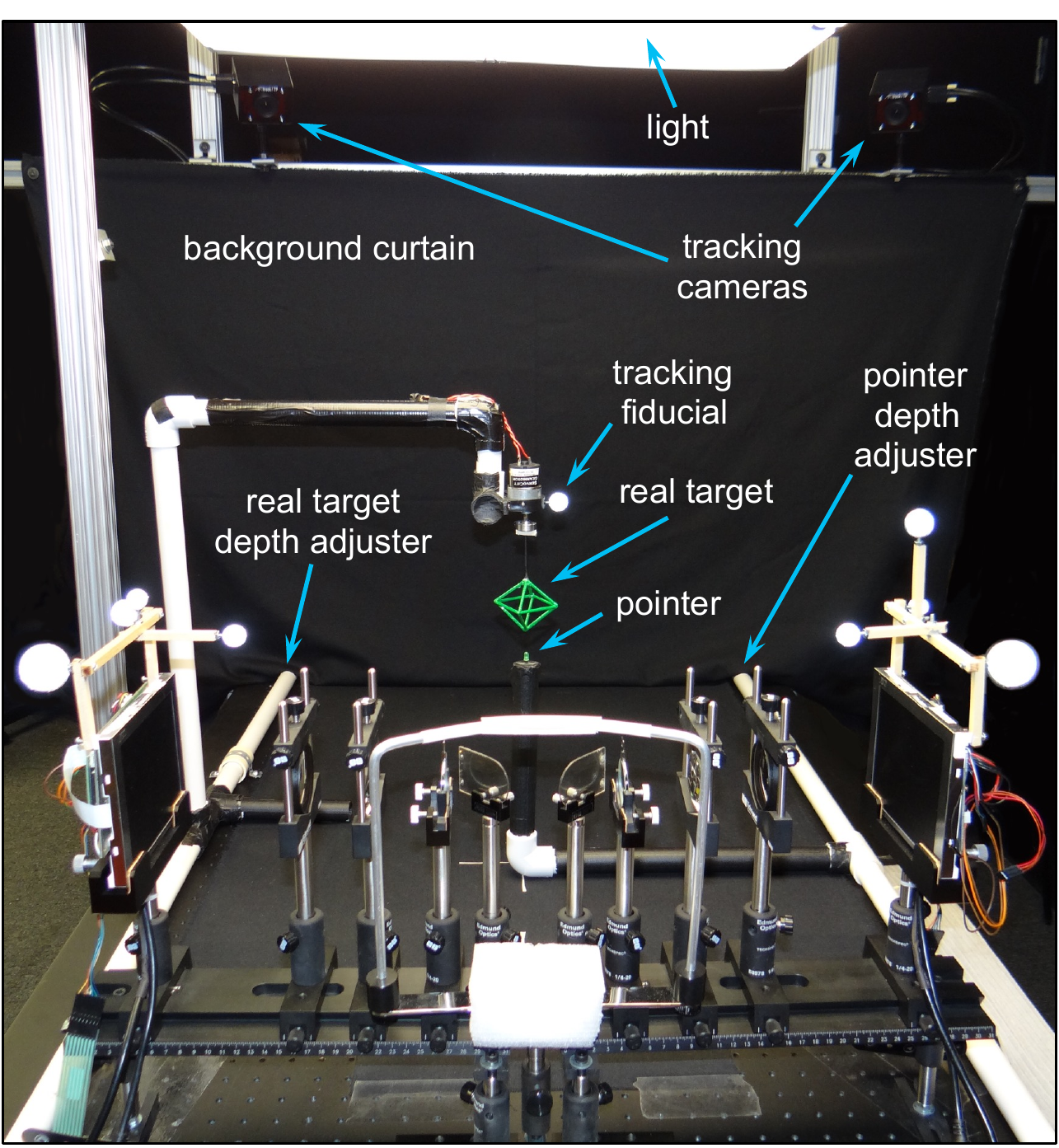}
\caption{The experimental setup.  The AR haploscope was mounted on the end of an optical breadboard.  Real and AR targets were positioned at different depths from the observer.  The depth of the targets was matched by changing the position of the pointer.}
\label{f:haplo-exp}
\end{figure}

Fig.~\ref{f:haplo-exp} shows the experimental setup.  The haploscope was mounted on the end of an optical breadboard, 244~cm long by 92~cm wide.  The breadboard was supported by a custom-built aluminum table, with six legs.  Mounted to the legs of the table were six hydraulic jacks, which could lift the entire table, so the surface could be adjusted to be between 104 and 134~cm above the ground.  This adjustability allowed the table to be comfortably positioned for observers of many different heights.  Aluminum arms extending above the table supported tracking cameras, as well as an overhead light (Fig.~\ref{f:haplo-exp}).  Because the tracking cameras and light were attached to the table, when the table height was adjusted, their distance above the table top remained constant.  Tracking was provided by a 2-camera TRACKPACK system, from A.R.T. GmbH. 

On both sides of the table, \emph{depth adjusters}---plastic pipes running through collars---could easily be slid back and forth in depth (Fig.~\ref{f:haplo-exp}).  When the real target was presented, it hung from an arm attached to the left-hand depth adjuster.  The real target was a wireframe octahedron, 5~cm wide by 6~cm high, constructed of balsa wood and painted green.  An electric motor rotated the target at 4 rpm.  Although slow, the rotation gave a definite sense of three-dimensional structure from motion, even when viewed monocularly.  The depth position of the real target was precisely measured by a tracking fiducial mounted to the arm (Fig.~\ref{f:haplo-exp}).  When an AR target was presented, the arm supporting the real target was removed.  The AR target was identical to the real target: a green octahedron that rotated at 4 rpm, rendered and viewed through the haploscope optics.  Only the green channel was used, which eliminated chromatic distortion.  Careful calibration ensured that the AR target matched the real target in size and position at all tested distances.  In addition, because accommodation lenses of different powers change the overall magnification of the optical system (Fig.~\ref{f:ray}), the calibration was repeated for every lens power.  The targets were located 29~cm above the tabletop, and seen against a black curtain hung 1.2 meters from the observer (Fig.~\ref{f:haplo-exp}).  The appearance of the real and AR targets was as similar as possible: the lighting and color of the real target made it appear to glow against an otherwise dark background, and it did not cast any visible shadows or reflections.  The table was covered with black cloth, which created a smooth and featureless surface under the target.

The matching task from Swan \etal~\cite{swan:2015} was replicated.  The pointer was made of green, translucent plastic, $\sim$4 mm in diameter, with a rounded top, mounted on an arm attached to the right-hand depth adjuster (Fig.~\ref{f:haplo-exp}).  Observers matched the target depth by sliding the depth adjuster until the pointer was directly below the bottom point of the rotating target.  The distance between the bottom of the target and the top of the pointer was $\sim$1~cm.  The depth position of the pointer was precisely measured by a tracking fiducial mounted on the arm (not visible in Fig.~\ref{f:haplo-exp}). 

\subsubsection {Experimental Design}

\ParLabelFirst{Observers:} 40 \emph{observers} were recruited from a population of university students and staff.  The observers ranged in age from 18 to 38; the mean age was 20.9, and 18 were male and 22 female.  10 observers were paid \$12 an hour, and the rest received course credit. 

\ParLabel{Independent Variables:} Observers saw 4 different \emph{conditions}: real consistent, AR collimated, AR consistent, and AR midpoint.  The target object appeared at 5 different \emph{distances} from the observer: 33.3, 36.4, 40, 44.4, and 50~cm, which correspond to 3, 2.75, 2.5, 2.25, and 2~D.  Observers saw 6 \emph{repetitions} of each distance. 

In the \emph{real consistent} condition, observers saw the real target object (Fig.~\ref{f:haplo-exp}), which, by definition, was always presented at a focal distance that was  consistent with its actual distance.  In the remaining conditions, the AR target was seen.  In the \emph{AR collimated} condition, a 0 D plain glass accommodation lens was used, presenting the target at optical infinity.  In the \emph{AR consistent} condition, the accommodation lens power---3, 2.75, 2.5, 2.25, or 2 D---was always consistent with the target's presented distance.  Finally, in the \emph{AR midpoint} condition, the 2.5 D accommodation lens was used, presenting the target at a focal distance of 40~cm.  

\ParLabel{Dependent Variables:} The primary dependent variable was \emph{judged distance}---the measured position of the pointer (Fig.~\ref{f:haplo-exp}).  
In addition, \emph{error} $=$ \emph{judged distance} $-$ \emph{actual distance} was also calculated (Fig.~\ref{f:swan2015}).  

\ParLabel{Design:} A mixed design was used, with condition varying between observers, and distance and repetition varying within each observer.  There were 10 observers in each condition, and the presentation order of condition varied in a round-robin fashion, so each group of 4 observers covered all conditions.  For each observer, distance $\times$ repetition was randomly permuted, with the restriction that the distance changed every trial.  Therefore, each observer completed $5\ \mbox{(distance)} \times 6\  \mbox{(repetition)} = 30\ \mbox{trials}$, and the experiment collected a total of $40\  \mbox{(observers)} \times 30\ \mbox{(trials)} = 1200\ \mbox{data points}$.

\subsubsection {Procedure}

After receiving an explanation of the experimental procedures, an observer gave informed consent.  Then, they took a stereo vision test, which measured their sensitivity to depth changes encoded by binocular disparity.  Next, the observer's inter-pupillary distance was measured, using a pupilometer set to optical infinity, and the haploscope was adjusted to match this distance.  The task was then explained, using the real target and the pointer.  If the observer indicated that, when working at the demonstrated distances, they would normally wear corrective optics (glasses or contacts), they were instructed to wear the optics.  Observers then donned safety goggles, which easily fit over glasses.  The googles had 3.5~cm circular openings for each eye, and were otherwise covered with black gaffer tape.  The size of these openings was calibrated so that, when looking through the haploscope optics, observers could see the complete field of view provided by the optical combiners, but their peripheral view of the rest of the haploscope was blocked.  The chinrest and forehead rest were adjusted so that the observer's eyes were approximately centered within the optical combiners, and the haploscope pivot points were approximately centered under the eyes' rotational centers (Figs.~\ref{f:haplo}, \ref{f:rotate}).  The table and chair heights were adjusted so the observer was sitting comfortably.

The observer then completed one of the four conditions.  The pointer was placed at a random position within the trackable distance of 23 to 67~cm from the observer, and the experimenter then displayed the first target distance.  Using their right hand to manipulate the pointer depth adjuster (Fig.~\ref{f:haplo-exp}), the observer moved the pointer from this starting position to match the target's depth.  The observer then closed their eyes, and the experimenter displayed the next target distance.  The observer then opened their eyes, and moved the pointer from the previously matched distance to the new distance.  This pattern continued until all trials were completed.  To display distances with the real target, the experimenter used the real target depth adjuster to slide the real target to the correct position.  For the AR target, the experimenter adjusted the angle of each haploscope arm, and swapped out the accommodation lenses as needed.  Regardless of condition, the procedures were as similar as possible, and the time required for each trial was approximately equal.  During real consistent trials, observers looked through the haploscope optics, even through the monitors were switched off.  

After the trials, the observer was debriefed.  The overall experiment took approximately one hour. 

\begin{figure*}[!t]
\centering
\includegraphics[width=1.5\FigWidth]{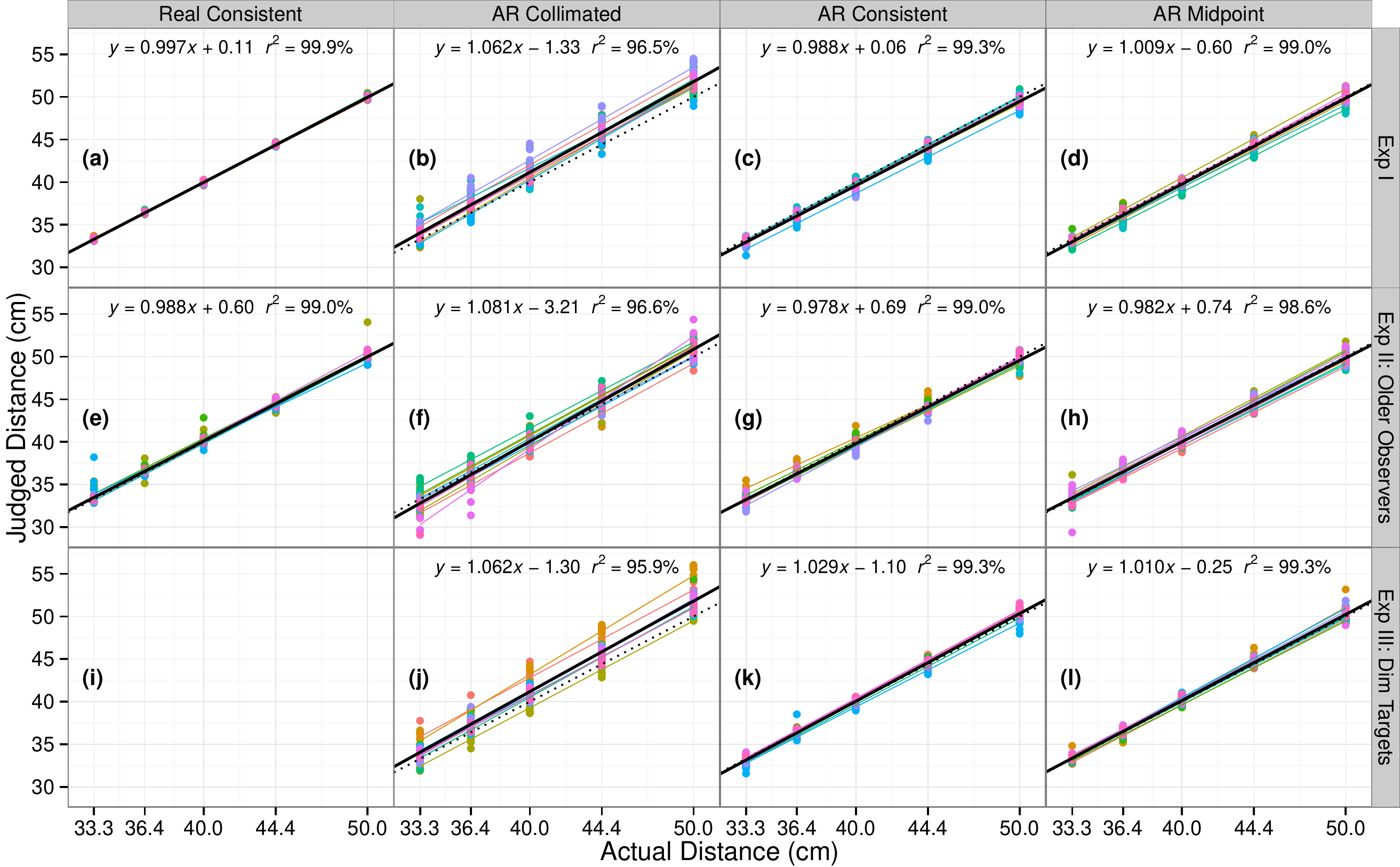} %
\caption{The results for all three experiments, plotted as judged against actual distance.  Each panel shows the individual data points, color coded according to observer, and fit with a color-coded regression line per observer.  For each panel, the dotted line represents veridical performance, and the solid black line is the overall regression line.  The corresponding regression equation is at the top of each panel.  The $r^{2}$ values indicate that the data in each panel is very well described by the regression equation.  A separate group of 10 observers provided the data for each panel.  The graph summarizes  $N = 1200$ (Experiments~I and II) and $N = 900$ (Experiment~III) data points.}
\label{f:jd}
\end{figure*}

\subsection {Analysis}

Similar to Swan \etal~\cite{swan:2015}, the data was analyzed by examining the slopes and intercepts of linear equations that predict judged distance from actual distance.  Multiple regression methods determine if the slopes and intercepts significantly differ (Pedhazur~\cite{pedhazur:1982}, Cohen \etal~\cite{cohen:2003}).  For data with this structure, multiple regression methods are preferable to ANOVA analysis, because multiple regression allows the prediction of a continuous dependent variable (judged distance) from a continuous independent variable (actual target distance), as well as a categorical independent variable (condition).  In contrast, ANOVA analysis only examines categorical independent variables, which results in a significant loss of power when an independent variable is inherently continuous (Pedhazur~\cite{pedhazur:1982}).  In addition, multiple regression yields slopes and intercepts, which as descriptive statistics are more useful than means, because they directly describe functions that predict judged distances from actual target distances.  Finally, multiple regression methods focus on effect size, as opposed to significance; an analytic approach advocated by many in the applied statistics community (Cohen \etal~\cite{cohen:2003}).\footnote{%
Custom analysis software, developed by the third author, was used, which implements methods described by Pedhazur~\cite{pedhazur:1982}.  A more detailed discussion of the application of multiple regression methods to depth perception data is available in Swan \etal~\cite{swan:2015}.}%

\begin{figure}[!t]
\centering
\includegraphics[width=1\FigWidth]{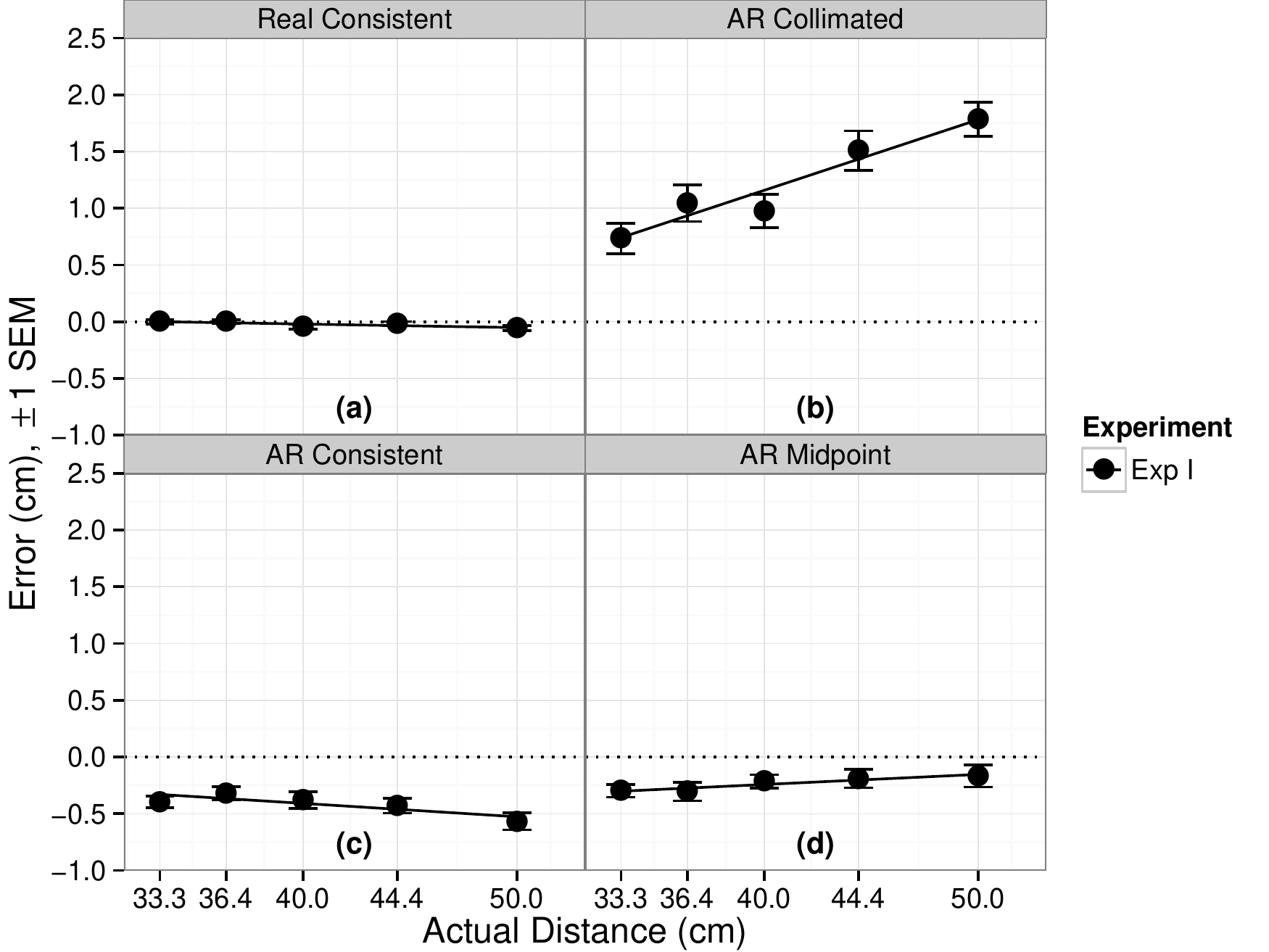}
\caption{Experiment I, plotted as mean error against distance $(N = 1200)$.} 
\label{f:e1-err}
\end{figure}

\begin{figure}[!t]
\centering
\includegraphics[width=1\FigWidth]{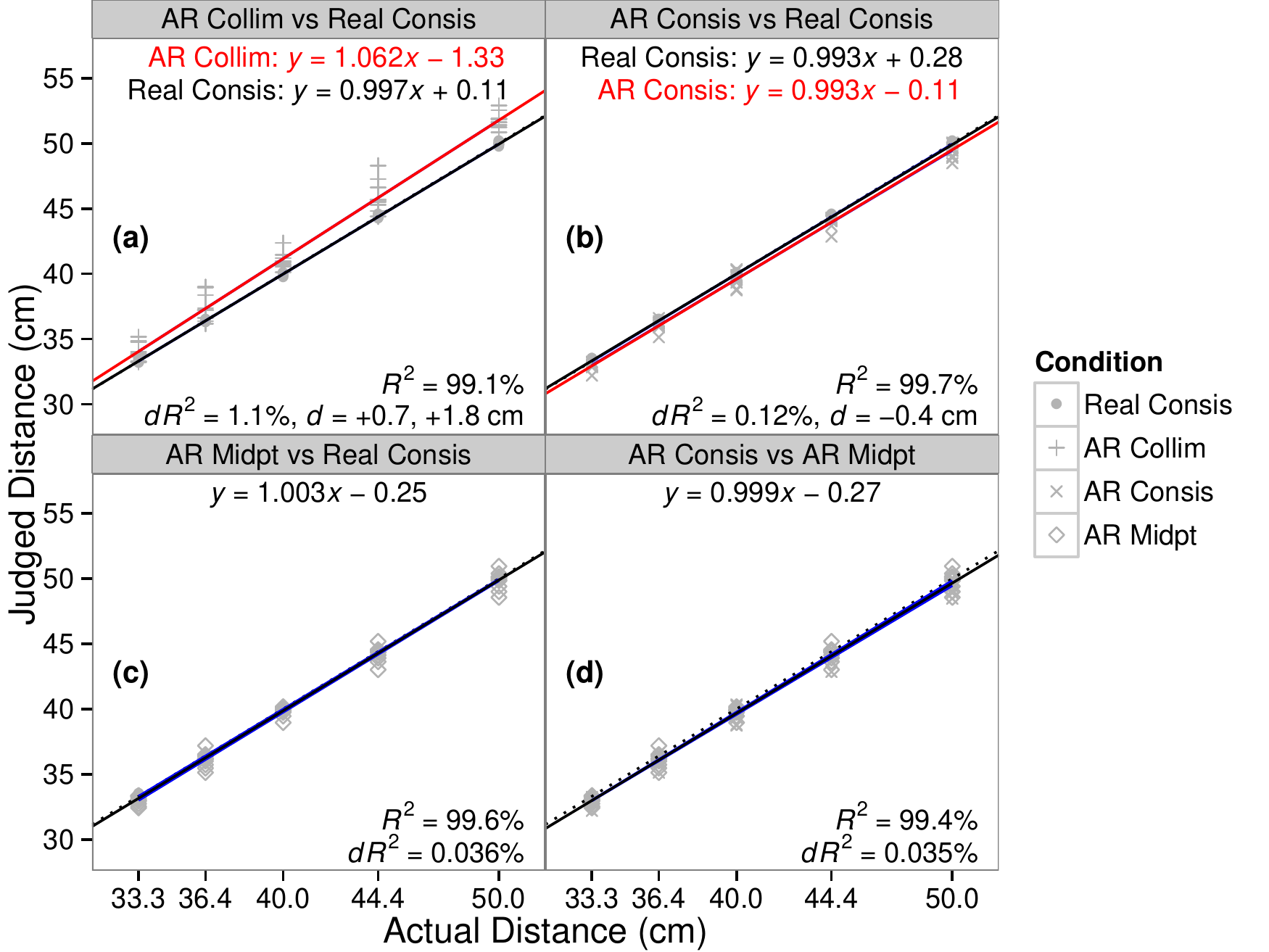}
\caption{Experiment~I, multiple regression analysis, plotted as a scatterplot of judged against actual distance, with $N = 200$ ghosted data points.  The thin dashed lines represent veridical performance.  Blue lines represent fitted regression equations from Fig.~\ref{f:jd}.  Black and red lines represent the linear regressions shown in each panel.  Blue lines are not visible when overlaid by black or red lines; the degree of blue line visibility is a graphical indication of how closely the regressions in each panel agree with the regressions from Fig.~\ref{f:jd}.} 
\label{f:e1-MR}
\end{figure}

Figs.~\ref{f:jd}a--d and \ref{f:e1-err} show the results from Experiment~I, plotted as a scatterplot of judged against actual distance (Fig.~\ref{f:jd}), as well as mean error against distance (Fig.~\ref{f:e1-err}).  Both figures indicate that the data is very well fit by linear regressions; note the $r^{2}$ values in Fig.~\ref{f:jd}.  Fig.~\ref{f:e1-MR} shows multiple regression analysis, which compares pairs of panels from Fig.~\ref{f:jd} against each other; each panel in Fig.~\ref{f:e1-MR} examines two independent variables: a continuous variable (actual distance), and a categorical variable (a pair of panels from Fig.~\ref{f:jd}).  To properly account for repeated measurements, for each observer at each distance, the responses were averaged over the 6 repetitions, reducing the size of the analyzed dataset from 1200 to 200 points---note the reduced density of points in Fig.~\ref{f:e1-MR} relative to Fig.~\ref{f:jd}a--d.

Each panel in Fig.~\ref{f:e1-MR} compares two regression equations from Fig.~\ref{f:jd}.
The multiple regression analysis operates in the following manner: 
First, the \emph{slopes} of the equations are tested to see if they significantly differ.  If they do, as in Fig.~\ref{f:e1-MR}a, both equations from Fig.~\ref{f:jd} are reported as the best overall description of the data in the panel.
If the slopes of the equations do not significantly differ, then the \emph{intercepts} of the equations are tested to see if they significantly differ.  This test first sets the slopes of the equations---which do not differ---to a common value.  If the intercepts significantly differ, as in Fig.~\ref{f:e1-MR}b, two regression equations, with slopes adjusted to a common value, are reported as the best overall description of the data in the panel.
If neither the slopes nor the intercepts significantly differ, as in Fig.~\ref{f:e1-MR}c, then the data from both panels is combined, and a regression over the combined data is reported as the best overall description of the data in the panel.  
Therefore, this multiple regression analysis yields three possible outcomes, which by chance are illustrated in the first three panels of Fig.~\ref{f:e1-MR}: (1) the slopes significantly differ (Fig.~\ref{f:e1-MR}a), (2) the slopes do not differ but the intercepts significantly differ (Fig.~\ref{f:e1-MR}b), or (3) neither the slopes nor the intercepts significantly differ (Fig.~\ref{f:e1-MR}c).  

In each case, the panel also indicates two measures of effect size: 
(1)~the overall $R^{2}$ value, the percentage of variation in the panel explained by the linear regressions, and 
(2)~$dR^{2}$, the percentage of variation explained by the change in the categorical variable.  If $dR^{2}$ is too small, hypothesis testing is not performed, because any statistical differences would be too small to be meaningful (Pedhazur \cite{pedhazur:1982}).  Based on the results reported in this paper, hypothesis testing is only conducted when  $dR^{2} \geq 0.08\%$.
Finally, for each panel, if there is a statistical difference in either slope or intercept, then the distance, $d$, in cm, between the fitted regression lines is also reported.  When there is a difference in slope, as in Fig.~\ref{f:e1-MR}a, $d$ is reported at the minimum and maximum $x$ values (33.3 and 50~cm).  When there is a difference in intercept, as in Fig.~\ref{f:e1-MR}b, then the regression lines are the same distance apart for every $x$, and only one $d$ value is reported.  $d$ is a signed value; $d > 0$ indicates a distance farther from the observer, and $d < 0$ closer to the observer.

\subsection {Results}

\ParDesFirst{Real consistent very accurate} Fig.~\ref{f:e1-err}a indicates that observers were extremely accurate in the real consistent condition.  The mean error is $-0.2$~mm, and the slope of the regression for Fig.~\ref{f:e1-err}a, $y = -0.003x + 0.11$, does not significantly differ from 0 ($F_{1,48} = 1.63, p = 0.21$).  Note that this is statistically equivalent to testing whether the slope in Fig.~\ref{f:jd}a differs from 1. 

\ParDes{AR collimated increasingly overestimated} When AR collimated is compared to real consistent (Fig.~\ref{f:e1-MR}a), the slopes significantly differ ($F_{1,96} = 10.7,  p = 0.001$), indicating that the AR collimated targets were overestimated, from $+$0.7 to $+$1.8~cm (Fig.~\ref{f:e1-err}b).

\ParDes{AR consistent underestimated} When AR consistent is compared to real consistent (Fig.~\ref{f:e1-MR}b), the slopes do not significantly differ ($F_{1,96} =  0.68, p = 0.41$), but the intercepts do ($F_{1,97} = 41.3,  p < 0.001$), indicating that the AR consistent targets were underestimated by a constant $-$0.4~cm (Fig.~\ref{f:e1-err}c).

\ParDes{AR midpoint equivalent to real consistent} When AR midpoint is  compared to real consistent (Fig.~\ref{f:e1-MR}c), the effect size of the difference is 0.036\% of the variation, which is too small for any statistical differences to be meaningful.  Therefore, the joint data is best fit by a single equation, indicating that AR midpoint targets were accurately matched (Fig.~\ref{f:e1-err}d). 

\ParDes{AR consistent and AR midpoint equivalent} When AR consistent is compared to AR midpoint (Fig.~\ref{f:e1-MR}d), the effect size is 0.035\%,  also too small for any statistical differences to be meaningful.  Therefore, matches of AR consistent and AR midpoint targets were equivalent (Fig.~\ref{f:e1-err}c, d).

\subsection {Discussion}

\begin{figure}[!t]
\centering
\includegraphics[width=1\FigWidth]{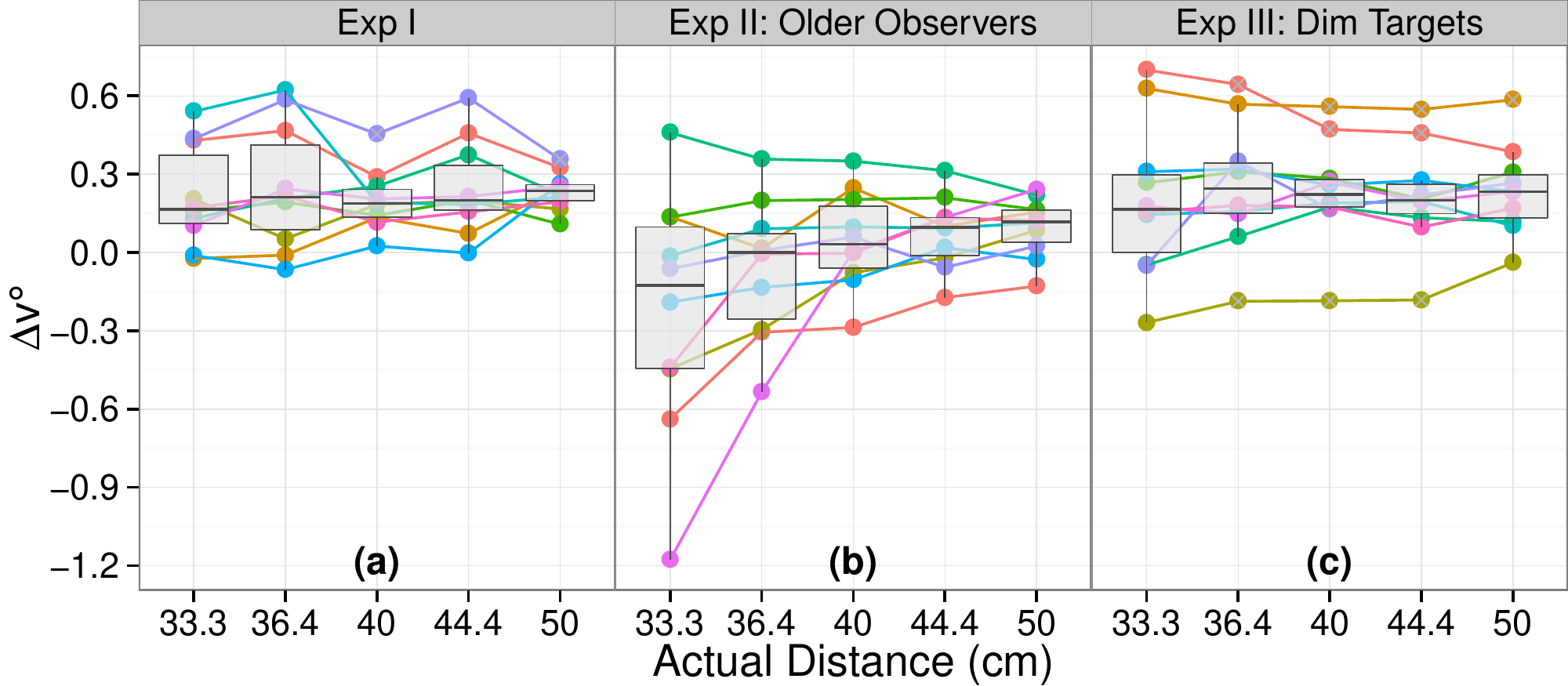}
\caption{For the \emph{AR collimated} condition, the change in vergence angle $\Delta v = \alpha - \beta$ (Fig.~\ref{f:model}), when an observer has matched the depth of the virtual target $\beta$ with the real pointer $\alpha$ (Fig.~\ref{f:haplo-exp}).  Each line in each panel is a different observer.  For all $N = 30$ observers, $\Delta v$ is approximately constant across all tested distances.  The boxplot gives the value for the median observer.}
\label{f:vdist}
\end{figure}

The first purpose (1) of Experiment~I was to replicate the real consistent and AR collimated conditions of Swan \etal~\cite{swan:2015} (Fig.~\ref{f:swan2015}).  The pattern in Figs.~\ref{f:e1-err}a, b indeed matches Fig.~\ref{f:swan2015}.  Given the many differences between the AR haploscope and the NVIS display used by Swan \etal~\cite{swan:2015}, this replication is consistent with the idea that this pattern of results generalizes to any collimated AR or stereo display.  
In addition, Swan \etal~\cite{swan:2015} hypothesized that collimation biases the eyes' vergence angle to rotate outwards by a constant amount (Fig.~\ref{f:model}).  
For each distance, Fig.~\ref{f:vdist}a shows $\Delta v$, the change in vergence angle,\footnote{%
$\Delta v = \alpha - \beta$, $\alpha = 2 \arctan(i/2x)$, and $\beta = 2 \arctan(i/2y)$, where $i$ is the observer's inter-pupillary distance, $x$ is the actual target distance, and $y$ is the judged distance (Fig.~\ref{f:jd}).  Note that using $x$ assumes that observers would match a real object with perfect accuracy, but the very accurate and precise results for the real consistent condition suggest this assumption is reasonable.} %
for the 10 AR collimated observers.  For all observers $\Delta v$ changes less than 0.5$^{\circ}$, and the median observer, seen in the boxplot, changes less than 0.072$^{\circ}$.  These small angular changes are consistent with the hypothesis that, within these reaching distances, the vergence angle bias is constant. 
 
The next purpose (2) was to test whether presenting AR objects at a focal distance that was \emph{consistent} with the distance specified by other depth cues, especially binocular disparity, would result in more accurate depth matches than what was seen in the AR collimated condition.  Figs.~\ref{f:e1-err}a, b, and c, as well as the analysis in Figs.~\ref{f:e1-MR}a and b, confirm this hypothesis: AR consistent is much more accurate than AR collimated, and for a consistent focal distance, real and AR targets do not differ in slope (Fig.~\ref{f:e1-MR}b).  

The final purpose (3) was to test whether presenting AR objects at a focal distance equal to the midpoint of the tested range would result in similar performance as the consistent condition.  Figs.~\ref{f:e1-err}c and d, as well as the analysis in Figs.~\ref{f:e1-MR}c and d, indicate that, when the focus was set to the midpoint, 
matching was indeed just as accurate.

\section {Experiment II: Age}
\label{s:exII}

As discussed in Section~\ref{s:age}, increasing age leads to presbyopia, a decline in the ability of the eyes to accommodate to different focal distances.  Experiment~I found significant negative effects of collimation, but all of the observers were young, with a mean age of 20.9, and therefore likely not presbyopic.  In addition, as discussed in Section~\ref{s:age}, although older people are worse than younger people at many perceptual tasks, recent studies have found that older people preserve their abilities in many tasks related to distance perception.  Therefore, it was unclear if older observers would replicate the effects observed in Experiment~I (Fig.~\ref{f:e1-err}).  Furthermore, this work was primarily inspired by medical AR applications, and the majority of medical professionals are old enough to suffer some degree of presbyopia.  Therefore, the purpose of Experiment~II was to replicate Experiment~I, using presbyopic observers, aged 40 and older.

\subsection {Method}

Other than the age of the observers, the methods of Experiment~II were identical to Experiment~I.  40 \emph{observers} were recruited from a population of university and community members.  The observers ranged in age from 41 to 80; the mean age was 55.6, and 19 were male and 21 female.  6 observers were paid \$10 an hour, 33 were paid \$12 hour, and one was not paid.  Each observer completed $5\ \mbox{(distance)} \times 6\  \mbox{(repetition)} = 30\ \mbox{trials}$, and the experiment collected a total of $40\  \mbox{(observers)} \times 30\ \mbox{(trials)} = 1200\ \mbox{data points}$.

\subsection {Results}

\begin{figure}[!t]
\centering
\includegraphics[width=1\FigWidth]{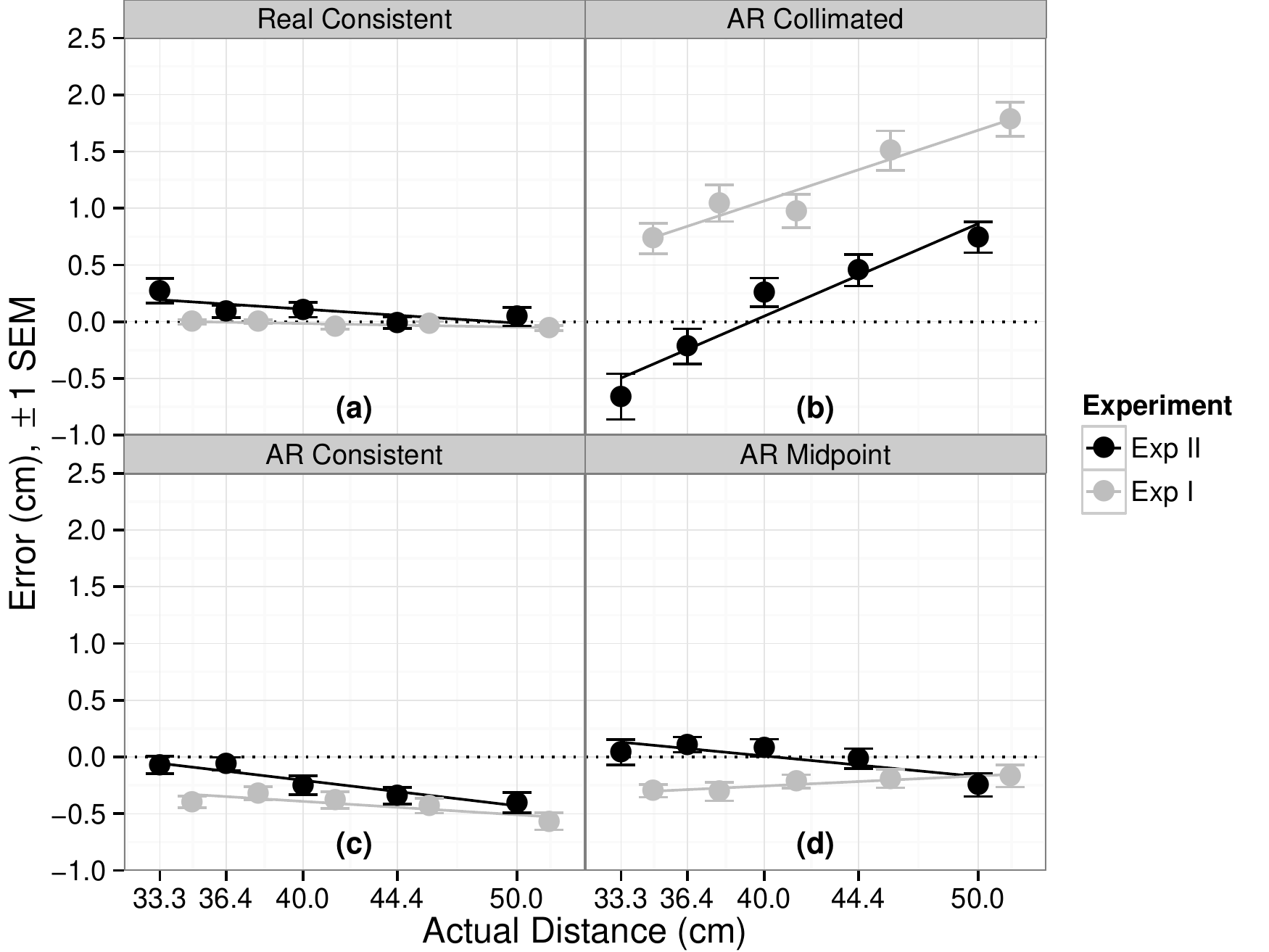}
\caption{Experiment~II: older observers, plotted as mean error against actual distance $(N = 1200)$.  For comparison, Experiment~I's results are also shown in light grey, offset along the $x$-axis for clarity.}
\label{f:e2-err}
\end{figure}

\begin{figure}[!t]
\centering
\includegraphics[width=1\FigWidth]{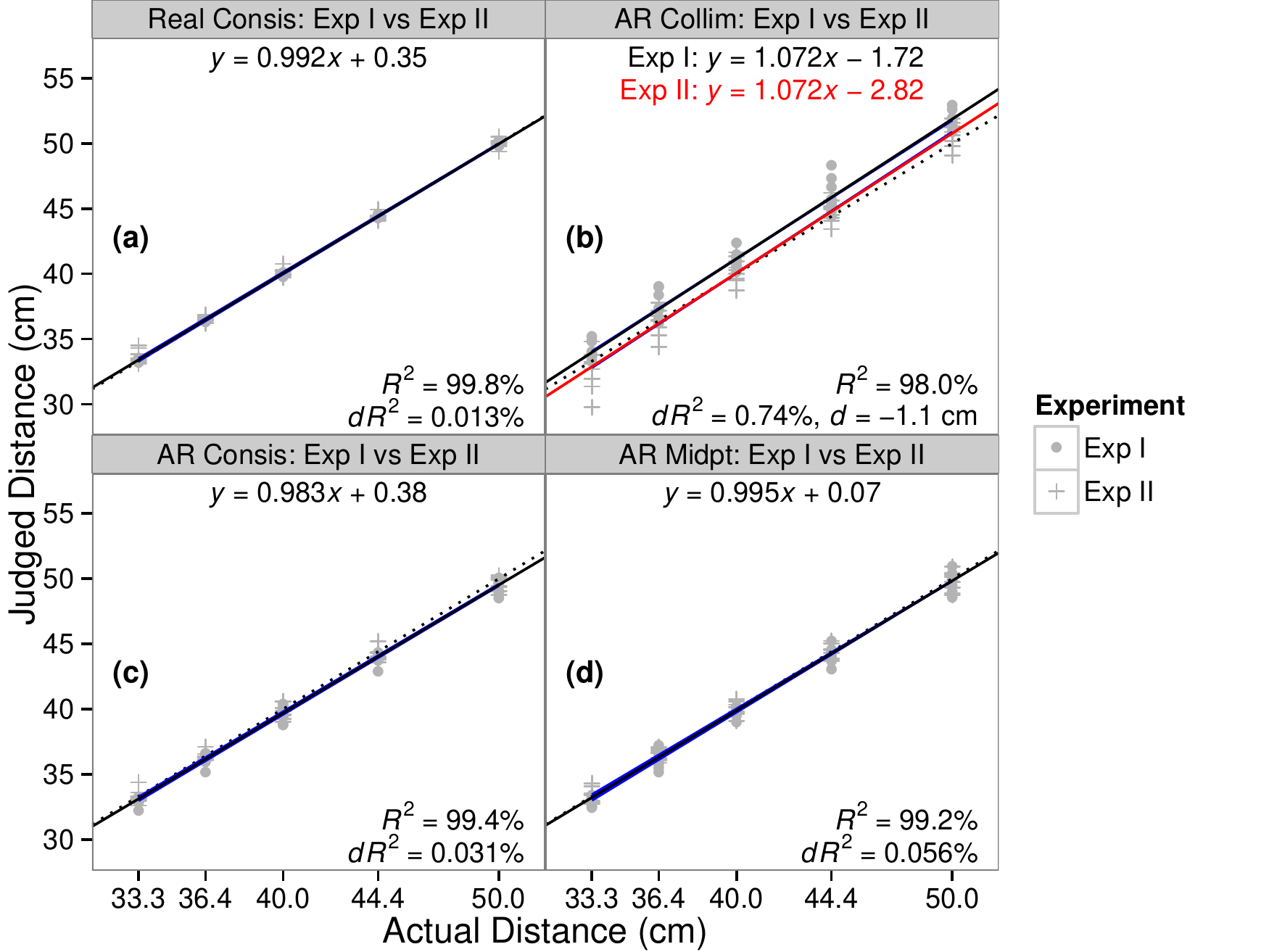}
\caption{Experiment~I versus II, the effect of age, multiple regression analysis, with $N = 400$ ghosted data points.  See the caption for Fig~\ref{f:e1-MR}.}
\label{f:e12-MR}
\end{figure}

\begin{figure}[!t]
\centering
\includegraphics[width=1\FigWidth]{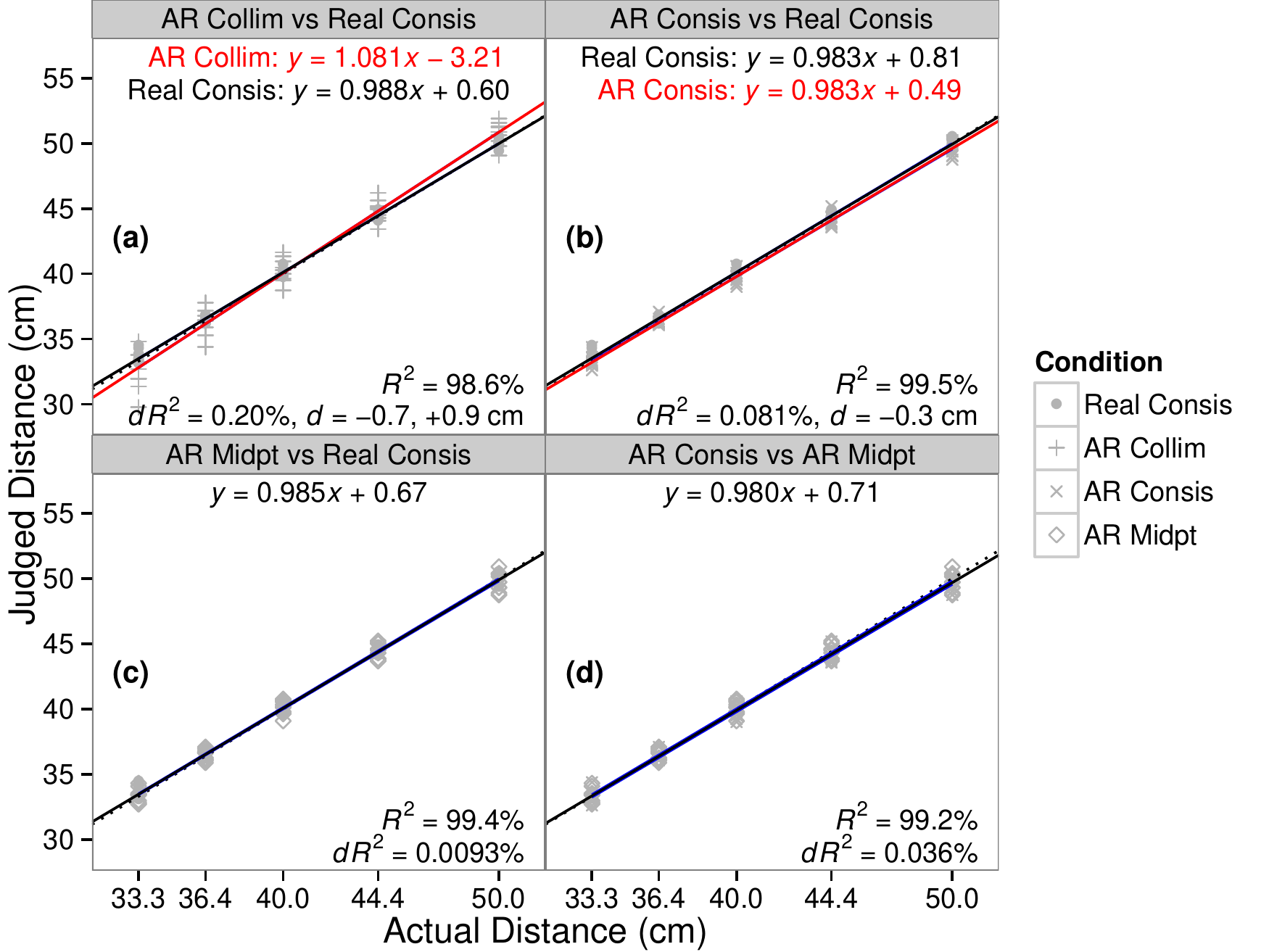}
\caption{Experiment~II, multiple regression analysis, older observers, with $N = 200$ ghosted data points.  See the caption for Fig~\ref{f:e1-MR}.}
\label{f:e2-MR}
\end{figure}

Fig.~\ref{f:jd}e--h shows the results from Experiment~II as scatterplots; the $r^{2}$ values indicate that the data continues to be very well fit by regression equations.  Fig.~\ref{f:e2-err} shows the results as error, with Experiment~I's results also shown for comparison.  Figs.~\ref{f:e12-MR} and \ref{f:e2-MR} show the results of multiple regression analysis. 

\ParDes{Older and younger only differ in AR collimated} Fig.~\ref{f:e12-MR} compares Experiment~I to Experiment~II condition by condition.
For the AR collimated condition (Fig.~\ref{f:e12-MR}b), the slopes do not significantly differ ($F_{1,96} =  0.38, p = 0.54$), but the intercepts do ($F_{1,97} = 35.7,  p < 0.001$); the older observers matched collimated AR targets a constant $-$1.1~cm closer to themselves than the younger observers (Fig.~\ref{f:e2-err}b).  For the remaining conditions, the effect size of the difference between Experiments~I and II, 0.013\% (Fig.~\ref{f:e12-MR}a), 0.031\% (Fig.~\ref{f:e12-MR}c), and 0.056\% (Fig.~\ref{f:e12-MR}d), is too small for any statistical differences to be meaningful.  Therefore, for the real consistent, AR consistent, and AR midpoint conditions, the results for the older observers and the younger observers are equivalent. 

\ParDes{Real consistent very accurate} Fig.~\ref{f:e2-err}a indicates that older observers were very accurate when matching the distance of real targets.  The mean error is $+0.4$ mm, and the slope of the linear model for Fig.~\ref{f:e2-err}a, $y = -0.012x + 0.60$, does not significantly differ from 0 ($F_{1,48} = 2.5, p = 0.12$).  Note that this is statistically equivalent to testing whether the slope in Fig.~\ref{f:jd}e differs from 1. 

\ParDes{AR collimated increasingly overestimated} For the older observers, when AR collimated is compared to real consistent (Fig.~\ref{f:e2-MR}a), the slopes significantly differ ($F_{1,96} =  13.8, p < 0.001$); the AR collimated errors ranged from $-$0.7 to $+$0.9~cm (Fig.~\ref{f:e2-err}b).

\ParDes{AR consistent underestimated} For the older observers, when AR consistent is compared to real consistent (Fig.~\ref{f:e2-MR}b), the slopes do not significantly differ ($F_{1,96} =  0.56, p = 0.46$), but the intercepts do ($F_{1,97} = 16.6,  p < 0.001$); the AR consistent targets were underestimated by a constant $-$0.3~cm (Fig.~\ref{f:e2-err}c).

\ParDes{AR midpoint equivalent to real consistent} For the older observers, when AR midpoint is compared to real consistent (Fig.~\ref{f:e2-MR}c), the effect size is 0.0093\%, which is too small for any statistical differences to be meaningful.  Therefore, the AR midpoint targets were accurately matched (Fig.~\ref{f:e2-err}d). 

\ParDes{AR consistent and AR midpoint equivalent} For the older observers, when AR consistent is compared to AR midpoint (Fig.~\ref{f:e2-MR}d), the effect size is 0.036\%,  also too small for any statistical differences to be meaningful.  Therefore, the matches of the AR consistent and AR midpoint targets were equivalent (Fig.~\ref{f:e2-err}c, d).

\subsection {Discussion}

The purpose of Experiment~II was to replicate Experiment~I, using older, presbyopic observers.  According to Duane~\cite{duane:1912}, the younger observers in Experiment~I had an expected near focus of $\sim$8.3~cm ($\sim$11.5~D), while for these older observers the expected near focus was $\sim$68~cm ($\sim$1.5~D).

Experiment~II's results only differ for the AR collimated condition.  For collimated targets, older observers showed less overestimation than younger observers, with matches shifted towards the observer by a constant $-$1.1~cm (Fig.~\ref{f:e12-MR}b).  Older observers had a mean error of $+$0.12~cm, while younger observers had a mean error of $+$1.2~cm (Fig.~\ref{f:e2-err}b), and therefore older observers were on average \emph{more} accurate than younger observers.  However, the slope, $b =$ 1.072, is the same for both sets of observers (Fig.~\ref{f:e12-MR}b), and differs significantly from the slope for the real consistent condition (Figs.~\ref{f:e1-MR}a and \ref{f:e2-MR}a).  Therefore, for both younger and older observers, matches of collimated targets were inaccurate, and increasingly overestimated with increasing distance.  In addition, for each distance, Fig.~\ref{f:vdist}b shows $\Delta v$, the change in vergence angle for the 10 older AR collimated observers.  For 9 of the 10 observers $\Delta v$ changes less than 0.6$^{\circ}$, for the outlying observer it changes 1.4$^{\circ}$, and the median observer changes less than 0.25$^{\circ}$.  These small angular changes are consistent with the hypothesis that, for both younger and older observers, the vergence angle bias  is constant.
 
For the other conditions, the observer' age---and therefore the observers' ability to accommodate to different focal demands---made no difference.  Older observers were just as accurate as younger observers in matching the distance to real targets, as well as to AR targets with both consistent and midpoint focal cues.  These results are consistent with previous work that has found that older observers preserve their abilities in many tasks related to distance perception (Bian and Andersen~\cite{bian:2013}).

\section {Experiment III: Brightness}
\label{s:exIII}

However, a conflicting finding from both experiments is that AR consistent was underestimated, while AR midpoint was accurate \emph{and} AR consistent was equivalent to AR midpoint.  This is true for both Experiment~I (Fig.~\ref{f:e1-MR}) and Experiment~II (Fig.~\ref{f:e2-MR}).  These conflicting statistical results are likely due to the small effect size of AR consistent's underestimation ($d = -$0.4 and $d = -$0.3~cm, respectively).  Nevertheless, the underestimation is statistically significant, and was replicated among 20 observers with widely varying ages. 

\begin{figure}[!t]
\centering
\includegraphics[width=0.9\FigWidth]{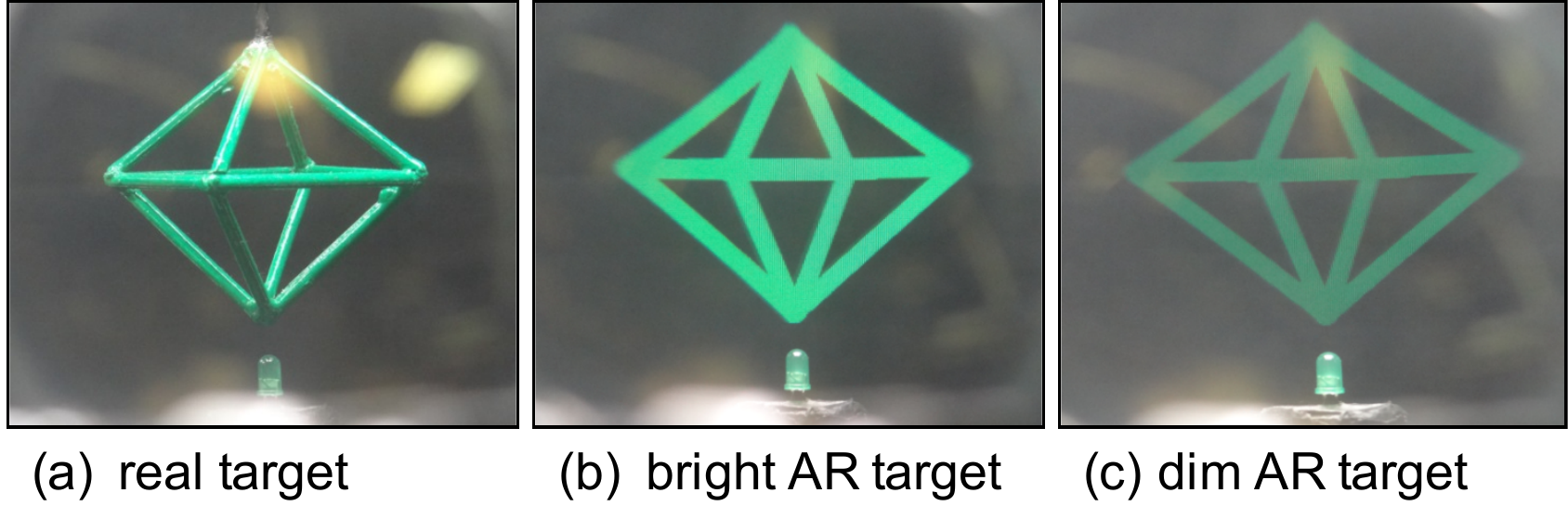}
\caption{Experiment~III examined target brightness.  (a) The real target.  (b) The bright AR target used in Experiments~I and II.  (c) The dim AR target used in Experiment III.}
\label{f:bright}
\end{figure}

As discussed in Section~\ref{s:bright}, brighter objects appear closer than similar-sized dimmer objects.  Figs.~\ref{f:bright}a and b show photographs, taken through the haploscope optics, of the real and AR targets used in Experiments~I and II.  The AR target appeared brighter than the real target.\footnote{%
Note that brightness is the perceptual experience of luminance, and cannot be directly measured or captured with a camera.  The luminance of the targets was measured (Singh~\cite{singh:2013}).} %
For Experiment~III, the brightness was reduced, until the AR and real targets appeared to have the same brightness (Fig.~\ref{f:bright}c).  The purpose of Experiment~III was to determine if the dim AR target would increase the accuracy of the AR consistent condition.  

\subsection {Method}

Other than the brightness of the AR target, the methods of Experiment~III were identical to Experiment~I.  Because the real target object did not change, that condition was not replicated.  To facilitate comparison with Experiment~I, younger observers were recruited, from a population of university students and staff.  The 30 \emph{observers} ranged in age from 17 to 24; the mean age was 19.8, and 21 were male and 9 female.  6 observers were paid \$12 an hour, and the rest received course credit.  Each observer completed $5\ \mbox{(distance)} \times 6\  \mbox{(repetition)} = 30\ \mbox{trials}$, and the experiment collected a total of $30\  \mbox{(observers)} \times 30\ \mbox{(trials)} = 900\ \mbox{data points}$.

\subsection {Results}

\begin{figure}[!t]
\centering
\includegraphics[width=1\FigWidth]{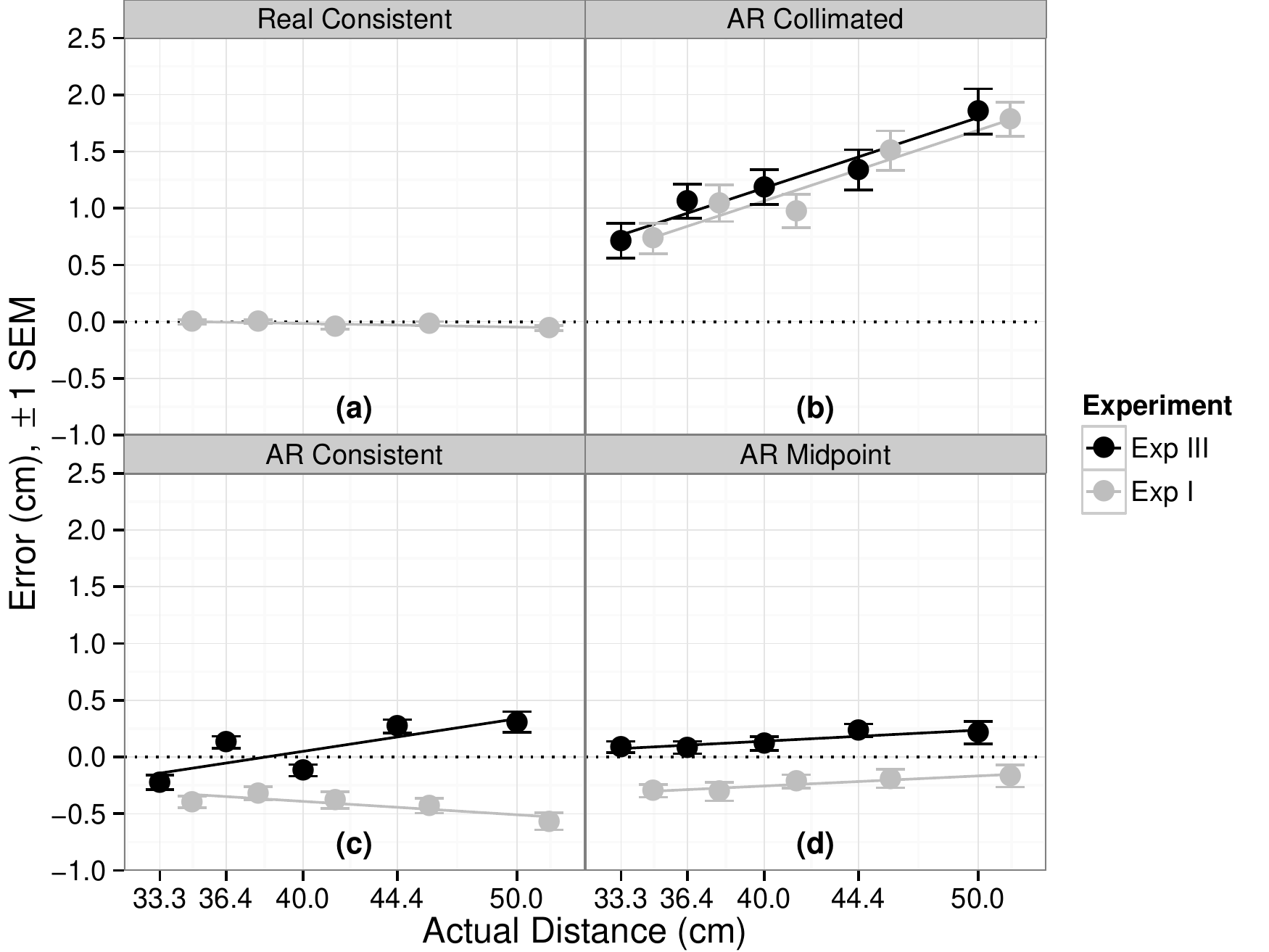}
\caption{Experiment~III: dim targets, plotted as mean error against actual distance $(N = 900)$.  For comparison, Experiment~I's results are also shown in light grey, offset along the $x$-axis for clarity.}
\label{f:e3-err}
\end{figure}

\begin{figure}[!t]
\centering
\includegraphics[width=1\FigWidth]{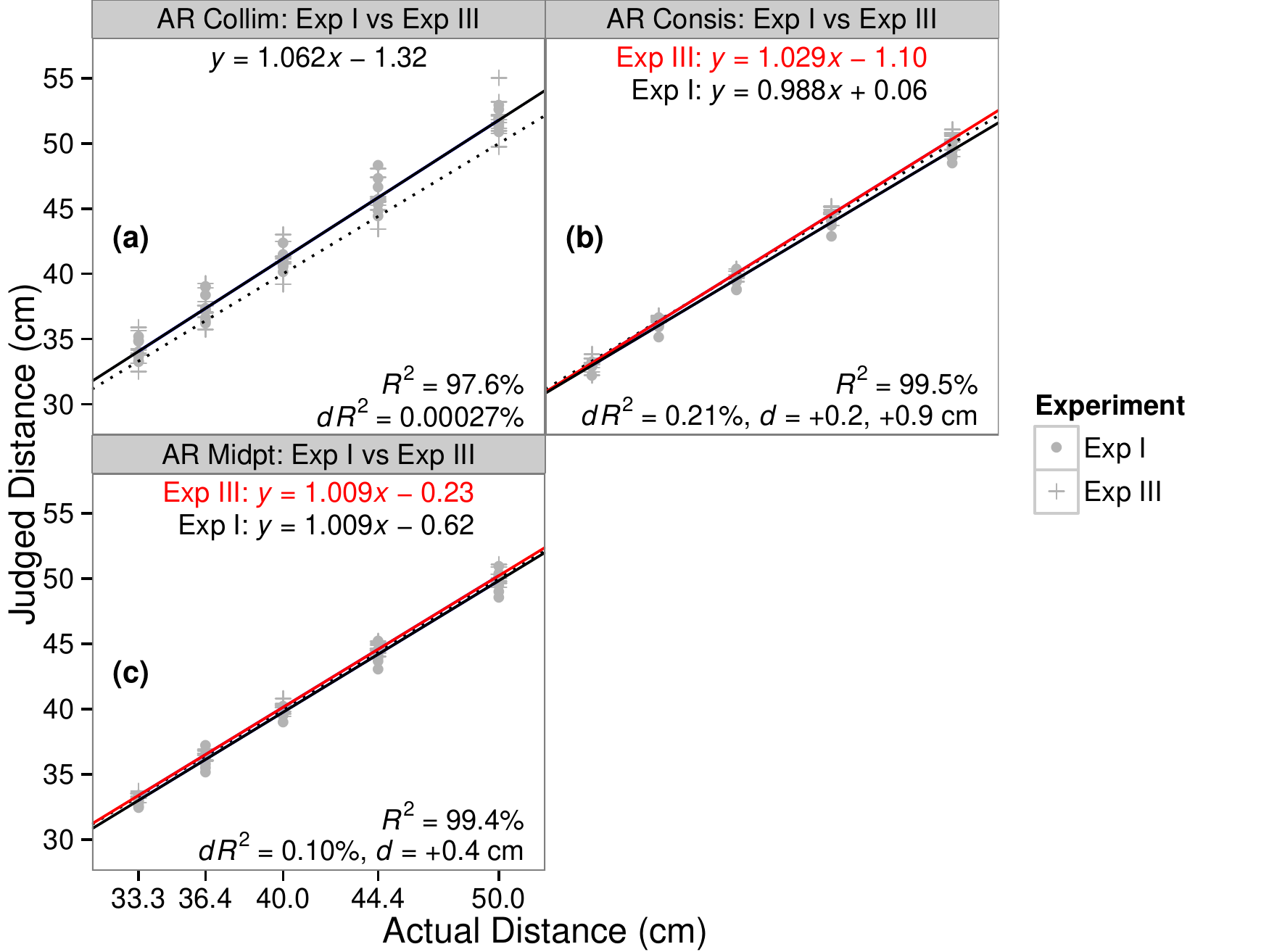}
\caption{Experiment~I versus III, the effect of brightness, multiple regression analysis, with $N = 300$ ghosted data points.  See the caption for Fig~\ref{f:e1-MR}.}  
\label{f:e13-MR}
\end{figure}

\begin{figure}[!t]
\centering
\includegraphics[width=1\FigWidth]{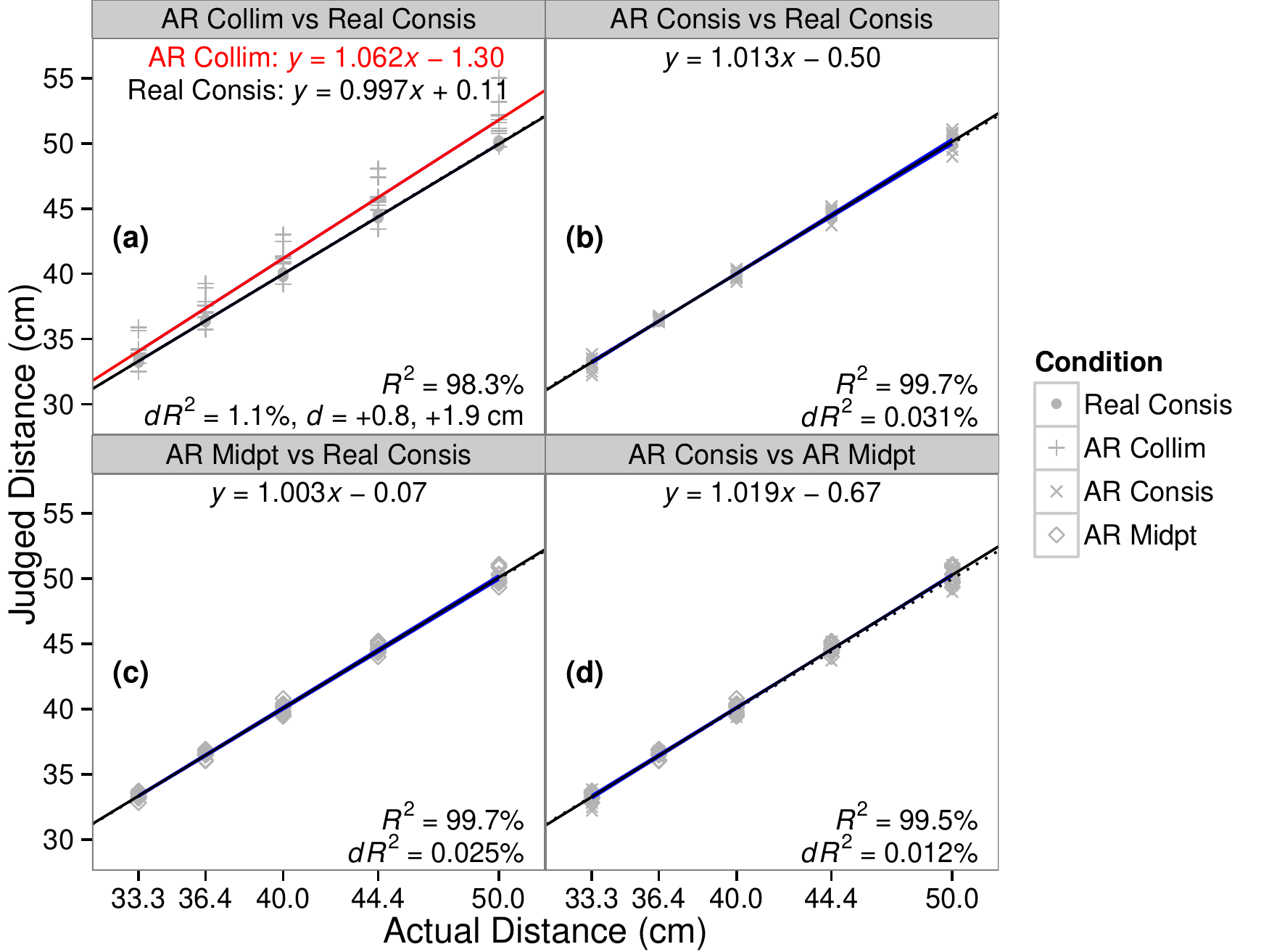}
\caption{Experiment~III, multiple regression analysis, dim targets, with $N = 200$ ghosted data points.  See the caption for Fig~\ref{f:e1-MR}.  The real consistent data is repeated from Experiment~I.}
\label{f:e3-MR}
\end{figure}

Fig.~\ref{f:jd}j--l shows the results from Experiment~III as scatterplots; the $r^{2}$ values indicate that the data continues to be very well fit by regression equations.  Fig.~\ref{f:e3-err} shows the same results as error, with Experiment~I's results also shown for comparison.  Figs.~\ref{f:e13-MR} and \ref{f:e3-MR} show the results of multiple regression analysis. 

\ParDes{Dim targets differ in AR consistent and AR midpoint} Fig.~\ref{f:e13-MR} compares Experiment~I to Experiment~III condition by condition.  For the AR collimated condition (Fig.~\ref{f:e13-MR}a), the effect size of the difference is 0.00027\%, much too small for any statistical differences to be meaningful.  Therefore, the results for the dim targets and the bright targets are equivalent (Fig.~\ref{f:e3-err}b).  For the AR consistent condition (Fig.~\ref{f:e13-MR}b), the slopes significantly differ ($F_{1,96} = 7.7,  p = 0.007$), and therefore the dim targets were matched $+$0.2 to $+$0.9~cm farther than the bright targets (Fig.~\ref{f:e3-err}c).  And finally, for the AR midpoint condition (Fig.~\ref{f:e13-MR}c), the slopes do not significantly differ ($F_{1,96} <  0.01, p = 0.96$), but the intercepts do ($F_{1,97} = 17.7,  p < 0.001$), and therefore the dim targets were matched $+$0.4~cm farther than the bright targets (Fig.~\ref{f:e3-err}d).

\ParDes{AR collimated increasingly overestimated} When dim AR collimated is compared to real consistent from Experiment~I (Fig.~\ref{f:e3-MR}a), the slopes significantly differ ($F_{1,96} = 5.5,  p = 0.021$), indicating that the dim AR collimated targets were overestimated from $+$0.8 to $+$1.9~cm (Fig.~\ref{f:e3-err}b).

\ParDes{AR consistent equivalent to real consistent} When dim AR consistent is  compared to real consistent from Experiment~I (Fig.~\ref{f:e3-MR}b), the effect size of the difference is 0.031\%, which is too small for any statistical differences to be meaningful.  Therefore, the dim AR consistent targets were accurately matched (Fig.~\ref{f:e3-err}c).

\ParDes{AR midpoint equivalent to real consistent} When dim AR midpoint is  compared to  real consistent from Experiment~I (Fig.~\ref{f:e3-MR}c), the effect size of the difference is 0.025\%, which is too small for any statistical differences to be meaningful.  Therefore, the dim AR midpoint targets were accurately matched (Fig.~\ref{f:e3-err}d).

\ParDes{AR consistent and AR midpoint equivalent} When dim AR consistent is compared to dim AR midpoint (Fig.~\ref{f:e3-MR}d), the effect size is 0.012\%,  also too small for any statistical differences to be meaningful.  Therefore, the matches of dim AR consistent and dim AR midpoint targets were equivalent (Fig.~\ref{f:e3-err}c, d).

\subsection {Discussion}

The purpose of Experiment~III was to determine if a dim AR target, which has the same apparent brightness as the real target (Fig.~\ref{f:bright}), would increase the accuracy of the AR consistent condition.  When comparing Experiment~III to Experiment~I, there was no difference in matching AR collimated targets, but increased accuracy for both AR consistent and AR midpoint targets (Fig.~\ref{f:e13-MR}).  In addition, when combined with the real consistent data from Experiment~I, matches for the dim AR target were overestimated in the AR collimated condition, but accurate in both the AR consistent and AR midpoint conditions (Figs.~\ref{f:e3-MR}).  Therefore, for the AR consistent and AR midpoint conditions, the dim AR targets were as accurately matched as the real targets (Fig.~\ref{f:e3-err}).

In addition, the pattern of AR consistent being underestimated, while AR midpoint was accurate \emph{and} AR consistent was equivalent to AR midpoint, occurred for both the younger observers of Experiment~I (Fig.~\ref{f:e1-MR}) and the older observers of Experiment~II (Fig.~\ref{f:e2-MR}).  While Experiment~III only tested younger observers, the results are consistent with the hypothesis that older observers would also accurately match the depth of dim AR consistent and dim AR midpoint targets.  

Finally, for the dim AR collimated targets, for each distance Fig.~\ref{f:vdist}c shows $\Delta v$, the change in vergence angle, for the 10 AR collimated observers.  For all observers $\Delta v$ changes less than 0.4$^{\circ}$, and the median observer changes less than 0.12$^{\circ}$.  These small angular changes are consistent with the hypothesis that, for the dim AR targets, the vergence angle bias is still constant.
 
\section {General Discussion}

\ParLabelFirst{Constant Vergence Angle Bias:} As discussed in Section~\ref{s:intro}, Swan \etal~\cite{swan:2015} found that the AR collimated condition caused overestimation that increased linearly with distance, and proposed that this was caused by the collimation biasing the eyes' vergence angle to rotate outwards by a constant amount.  All of the experiments reported here replicated this result, and strongly support this hypothesis.  These findings are also consistent with the prediction, by Mon-Williams and Tresilian~\cite{monwilliams:2000}, that an inconsistent accommodative cue would bias perceived depth in the same direction as the accommodative cue (Fig.~\ref{f:verg-acc-con}).  However, Swan \etal~\cite{swan:2015} did not measure this vergence angle change, and it was not measured here.  In a future experiment, it should be directly measured. 

\ParLabel{Dim Targets:} The experiments found the most accurate matches for dim AR targets, which more closely matched the brightness of the real targets.  These results are consistent with previous work that finds brighter objects appear closer than dimmer objects (Ashley~\cite{ashley:1898}, Farn\`{e}~\cite{farne:1977}, Coules~\cite{coules:1955}).  However, it is interesting that the error for matching the dim AR targets disappeared, even though the error was calculated \emph{between-subjects}: in all of the experiments, different groups of 10 observers saw the real targets, the bright AR targets, and the dim AR targets.  It would have been less surprising to have found these errors in a within-subjects design, where observers made a judgment about two targets with different brightnesses, viewed simultaneously (e.g., Ashley~\cite{ashley:1898}, Farn\`{e}~\cite{farne:1977}, Coules~\cite{coules:1955}).  The errors may be related to the fact that AR targets are drawn with impoverished depth cues, and therefore brightness could be directly biasing the vergence angle.  If this hypothesis is true, it would be another component of accurate depth presentation that must be considered by AR practitioners.  A future experiment should examine whether the brightness of an AR object directly influences vergence angle. 

\ParLabel{Midpoint Accommodative Stimulus:} While the AR consistent condition was accurate for the dim AR targets, the AR midpoint condition was accurate across all of the experiments.  It is not clear why AR midpoint was accurate at both brightness levels, while AR consistent was not.  
Nevertheless, the practical implication is that, because the AR midpoint condition was at least as accurate as the AR consistent condition, for AR applications requiring accurate near-field depth matching, it is sufficient for the focal demand to be set to the middle of the working volume.  

However, the positive results for the AR midpoint condition suggest comparison with light-field displays, which can simultaneously present multiple virtual objects at different focal distances.  In addition to solving the vergence-accommodation conflict, light-field displays are predicted to eventually become the dominant technology for all kinds of 3D experiences (Balram~\cite{balram:2014}).  However, although the technology is rapidly developing, AR light-field displays face many fundamental challenges and design tradeoffs, in areas such as  depth range, color resolution, spatial resolution, computational demands, and data throughput requirements (Wu \etal~\cite{wu:2014}).  Therefore, the AR midpoint results suggest that the level of engineering complexity required for these kinds of displays may not be necessary, especially for AR applications where the most important perceptual task is accurate matching at near-field distances (e.g., Edwards~\cite{edwards:2000}, Krempien \etal~\cite{krempien:2008}).

\ParLabel{Future Work:} As discussed in this section, errors detected in these experiments are likely due to vergence angle biases.  Therefore, useful future work would replicate these experiments while measuring vergence angle.  Possible methods for making this measurement  include binocular eye tracking (Wang \etal~\cite{duchowski:2014}), or nonius line methods (Ellis and Menges~\cite{ellis:1998}).  

In addition, because the AR haploscope was mounted to a tabletop, these experiments could not examine the depth cue of motion perspective.  Although some AR applications, such as the operating microscope described by Edwards \etal~\cite{edwards:2000}, are also mounted and therefore lack motion perspective, it is a very salient depth cue (Nagata \cite{nagata:1991}, Cutting and Vishton \cite{cutting:1995}), and should be examined in future experiments.  A head-mounted AR haploscope, such as the one used by McCandless \etal~\cite{mccandless:2000}, would allow a replication of these experiments that included motion perspective.

\section {Practical Implications}

For accurate near-field depth matching, the experiments reported here have the following implications:

\ParLabel{\textbullet} Collimated graphics should not be used.  A focal distance set to the middle of the depth range is a good as a focal distance optimized for every virtual object.

\ParLabel{\textbullet} The brightness of virtual objects needs to match the brightness of real objects.

\ParLabel{\textbullet} Observers old enough to suffer age-related reductions in accommodative ability are just as accurate as younger observers. 

\ifCLASSOPTIONcompsoc
  \section*{Acknowledgments}
\else
  \section*{Acknowledgment}
\fi

This material is based upon work supported by the National Science Foundation, under awards IIS-0713609, IIS-1018413, and IIS-1320909, to J. E. Swan II.

\ifCLASSOPTIONcaptionsoff
  \newpage
\fi

\bibliographystyle{IEEEtran}
\bibliography{swan-ellis-singh-arXiv-v1}

\begin{thebibliography}{10}
\providecommand{\url}[1]{#1}
\csname url@samestyle\endcsname
\providecommand{\newblock}{\relax}
\providecommand{\bibinfo}[2]{#2}
\providecommand{\BIBentrySTDinterwordspacing}{\spaceskip=0pt\relax}
\providecommand{\BIBentryALTinterwordstretchfactor}{4}
\providecommand{\BIBentryALTinterwordspacing}{\spaceskip=\fontdimen2\font plus
\BIBentryALTinterwordstretchfactor\fontdimen3\font minus
  \fontdimen4\font\relax}
\providecommand{\BIBforeignlanguage}[2]{{%
\expandafter\ifx\csname l@#1\endcsname\relax
\typeout{** WARNING: IEEEtran.bst: No hyphenation pattern has been}%
\typeout{** loaded for the language `#1'. Using the pattern for}%
\typeout{** the default language instead.}%
\else
\language=\csname l@#1\endcsname
\fi
#2}}
\providecommand{\BIBdecl}{\relax}
\BIBdecl

\bibitem{kersten-oertel:2013}
M.~Kersten-Oertel, P.~Jannin, and D.~L. Collins, ``{The State of the Art of
  Visualization in Mixed Reality Image Guided Surgery},'' \emph{Comput. Medical
  Imaging and Graphics}, vol.~37, no.~2, pp. 98--112, 2013.

\bibitem{curtis:1998}
D.~Curtis, D.~Mizell, P.~Gruenbaum, and A.~Janin, ``Several devils in the
  details: Making an {AR} application work in the airplane factory,'' in
  \emph{Proc. of Intern. Workshop on Augmented Reality (IWAR)}, 1998, pp.
  47--60.

\bibitem{henderson:2009}
S.~J. Henderson and S.~Feiner, ``Evaluating the benefits of augmented reality
  for task localization in maintenance of an armored personnel carrier
  turret,'' in \emph{Intern. Symp. on Mixed and Augmented Reality (ISMAR)},
  2009, pp. 135--144.

\bibitem{edwards:2000}
P.~J. Edwards, A.~P. King, C.~R. Maurer~Jr., D.~A. de~Cunha, D.~J. Hawkes,
  D.~L.~G. Hill, R.~P. Gaston, M.~R. Fenlon, A.~Jusczyzck, A.~J. Strong, C.~L.
  Chandler, and M.~J. Gleeson, ``Design and evaluation of a system for
  microscope-assisted guided interventions {(MAGI)},'' \emph{IEEE Trans. on
  Medical Imaging}, vol.~19, no.~11, pp. 1082--1093, 2000.

\bibitem{krempien:2008}
R.~Krempien, H.~Hoppe, L.~Kahrs, S.~Daeuber, O.~Schorr, G.~Eggers, M.~Bischof,
  M.~W. Munter, J.~Debus, and W.~Harms, ``{Projector-Based Augmented Reality
  for Intuitive Intraoperative Guidance in Image-Guided 3D Interstitial
  Brachytherapy},'' \emph{International Journal of Radiation Oncology
  {\textbullet} Biology {\textbullet} Physics}, vol.~70, no.~3, pp. 944--952,
  Mar. 2008.

\bibitem{swan:2015}
J.~E. {Swan~II}, G.~Singh, and S.~R. Ellis, ``Matching and reaching depth
  judgments with real and augmented reality targets,'' \emph{IEEE Trans. on
  Visualization and Computer Graphics}, vol.~21, no.~11, pp. 1289--1298, 2015.

\bibitem{gabbard:2017}
J.~L. Gabbard, D.~G. Mehra, and J.~E. {Swan~II}, ``Effects of {AR} display
  context switching and focal distance switching on human performance,''
  \emph{article in submission}, 2017.

\bibitem{singh:2013}
G.~Singh, ``{Near-Field Depth Perception in Optical See-Through Augmented
  Reality},'' Ph.D. dissertation, Mississippi State University, Aug. 2013.

\bibitem{cutting:1995}
J.~E. Cutting and P.~M. Vishton, ``Perceiving layout and knowing distances: The
  integration, relative potency, and contextual use of different information
  about depth,'' in \emph{Handbook of Perception and Cognition: Perception of
  Space and Motion}, W.~Epstein and S.~Rogers, Eds.\hskip 1em plus 0.5em minus
  0.4em\relax Academic Press, 1995, vol.~5, pp. 69--117.

\bibitem{nagata:1991}
S.~Nagata, ``{How to Reinforce Perception of Depth in Single Two-Dimensional
  Pictures},'' in \emph{Pictorial Communication in Virtual and Real
  Environments}, S.~R. Ellis, M.~K. Kaiser, and A.~J. Grunwald, Eds.\hskip 1em
  plus 0.5em minus 0.4em\relax London: Taylor \& Francis, 1989, pp. 527--545.

\bibitem{gillam:1995}
B.~Gillam, ``The perception of space layout,'' in \emph{Perception of Space and
  Motion}, 2nd~ed., W.~Epstein and S.~Rogers, Eds.\hskip 1em plus 0.5em minus
  0.4em\relax Academic Press, 1995, pp. 23--67.

\bibitem{bingham:1998}
G.~P. Bingham and C.~C. Pagano, ``The necessity of a perception-action approach
  to definite distance perception: Monocular distance perception to guide
  reaching,'' \emph{Journal of Experimental Psychology: Human Perception and
  Performance}, vol.~24, no.~1, pp. 145--168, 1998.

\bibitem{monwilliams:1999}
M.~Mon-Williams and J.~R. Tresilian, ``Some recent studies on the extraretinal
  contribution to distance perception,'' \emph{Perception}, pp. 167--181, 1999.

\bibitem{landy:1995}
M.~S. Landy, L.~T. Maloney, E.~B. Johnston, and M.~Young, ``Measurement and
  modeling of depth cue combination: In defense of weak fusion,'' \emph{Vision
  Research}, vol.~35, no.~3, pp. 389--412, Feb 1995.

\bibitem{leigh:2015}
R.~J. Leigh and D.~S. Zee, \emph{The Neurology of Eye Movements}, 5th~ed.\hskip
  1em plus 0.5em minus 0.4em\relax Oxford University Press, USA, 2015.

\bibitem{semmlow:1983}
J.~L. Semmlow and G.~K. Hung, ``The near response: Theories of control,'' in
  \emph{Vergence Eye Movements: Basic \& Clinical Aspects}, C.~M. Schor and
  K.~J. Ciuffreda, Eds.\hskip 1em plus 0.5em minus 0.4em\relax Butterworth,
  1983, pp. 175--195.

\bibitem{ripps:1962}
H.~Ripps, N.~B. Chin, I.~M. Siegel, and G.~M. Breinin, ``The effect of pupil
  size on accommodation, convergence, and the ac/a ratio.'' \emph{Investigative
  Ophthalmology}, vol.~1, pp. 127--135, Feb 1962.

\bibitem{kersten:1983}
D.~Kersten and G.~E. Legge, ``Convergence accommodation,'' \emph{Journal of the
  Optical Society of America}, vol.~73, no.~3, pp. 332--338, Mar 1983.

\bibitem{monwilliams:2000}
M.~Mon-Williams and J.~R. Tresilian, ``Ordinal depth information from
  accommodation?'' \emph{Ergonomics}, vol.~43, no.~3, pp. 391--404, 2000.

\bibitem{kruijff:2010}
E.~Kruijff, J.~E. {Swan~II}, and S.~Feiner, ``{Perceptual Issues in Augmented
  Reality Revisited},'' in \emph{Intern. Symp. on Mixed and Augmented Reality
  (ISMAR)}, J.~Park, V.~Lepetit, and T.~H{\"o}llerer, Eds.\hskip 1em plus 0.5em
  minus 0.4em\relax Piscataway, NJ, USA: IEEE, 2010, pp. 3--12.

\bibitem{lambooij:2009}
M.~Lambooij, M.~Fortuin, I.~Heynderickx, and W.~IJsselsteijn, ``Visual
  discomfort and visual fatigue of stereoscopic displays: A review,''
  \emph{Journal of Imaging Science and Technology}, vol.~53, no.~3, pp.
  030\,201--030\,201--14, 2009.

\bibitem{hoffman:2008}
D.~M. Hoffman, A.~R. Girshick, K.~Akeley, and M.~S. Banks,
  ``Vergence--accommodation conflicts hinder visual performance and cause
  visual fatigue,'' \emph{Journal of Vision}, vol. 8(3), no.~33, pp. 1--30,
  2008.

\bibitem{swenson:1932}
H.~A. Swenson, ``{the Relative Influence of Accommodation and Convergence in
  the Judgment of Distance},'' \emph{Journal of General Psychology}, vol.~7,
  no.~2, pp. 360--380, 1932.

\bibitem{brenner:1998}
E.~Brenner and J.~M. Van~Damme, ``Judging distance from ocular convergence,''
  \emph{Vision Research}, vol.~38, no.~4, pp. 493--498, 1998.

\bibitem{owens:1980}
D.~A. Owens and H.~W. Liebowitz, ``Accommodation, convergence, and distance
  perception in low illumination,'' \emph{Optometry \& Vision Science},
  vol.~57, no.~9, pp. 540--550, 1980.

\bibitem{tresilian:1999}
J.~R. Tresilian, M.~Mon-Williams, and B.~M. Kelly, ``Increasing confidence in
  vergence as a cue to distance,'' \emph{Proc. of the Royal Society of London
  B}, vol. 266, pp. 39--44, 1999.

\bibitem{viguier:2001}
A.~Viguier, G.~Clément, and Y.~Trotter, ``Distance perception within near
  visual space,'' \emph{Perception}, vol.~30, pp. 115--124, 2001.

\bibitem{foley:1980}
J.~M. Foley, ``Binocular distance perception,'' \emph{Psychological Review},
  vol.~87, no.~5, pp. 411--434, 1980.

\bibitem{iavecchia:1988}
J.~H. Iavecchia, H.~P. Iavecchia, and S.~N. Roscoe, ``Eye accommodation to
  head-up virtual images,'' \emph{Human Factors}, vol.~30, no.~6, pp. 703--712,
  1988.

\bibitem{roscoe:1985}
S.~N. Roscoe, ``{Bigness Is in the Eye of the Beholder},'' \emph{Human
  Factors}, vol.~27, no.~6, pp. 615--636, 1985.

\bibitem{duane:1912}
A.~Duane, ``Normal values of the accommodation at all ages,'' \emph{Journal of
  the American Medical Association}, vol. LIX, pp. 1010--1013, 1912.

\bibitem{kasthurirangan:2006}
S.~Kasthurirangan and A.~Glasser, ``Age related changes in accommodative
  dynamics in humans,'' \emph{Vision Research}, vol.~46, pp. 1507--1519, Apr
  2006.

\bibitem{bian:2013}
Z.~Bian and G.~J. Andersen, ``Aging and the perception of egocentric
  distance,'' \emph{Psychology and Aging}, vol.~28, no.~3, pp. 813--825, 2013.

\bibitem{heron:2001}
G.~Heron, W.~N. Charman, and C.~M. Schor, ``Age changes in the interactions
  between the accommodation and vergence systems,'' \emph{Optometry and Vision
  Science}, vol.~78, no.~10, pp. 754--762, 2001.

\bibitem{andersen:1998}
G.~J. Andersen, A.~Saidpour, and M.~L. Braunstein, ``{Effects of Collimation on
  Perceived Layout in 3-D Scenes},'' \emph{Perception}, vol.~27, no.~11, pp.
  1305--1315, 1998.

\bibitem{watt:2005}
S.~J. Watt, K.~Akeley, M.~O. Ernst, and M.~S. Banks, ``{Focus Cues Affect
  Perceived Depth},'' \emph{Journal of Vision}, vol.~5, no.~10, pp. 834--862,
  2005.

\bibitem{mccurdy:1938}
E.~McCurdy, \emph{The Notebooks of Leonardo da Vinci, Volume II}.\hskip 1em
  plus 0.5em minus 0.4em\relax Reynal \& Hitchcock, 1938.

\bibitem{ashley:1898}
M.~L. Ashley, ``Concerning the significance of intensity of light in visual
  estimates of depth,'' \emph{Psychological Review}, vol.~5, no.~6, pp.
  595--615, 1898.

\bibitem{farne:1977}
M.~Farn\`{e}, ``Brightness as an indicator to distance: Relative brightness per
  se or contrast with the background?'' \emph{Perception}, vol.~6, pp.
  287--293, 1977.

\bibitem{coules:1955}
J.~Coules, ``Effect of photometric brightness on judgments of distance,''
  \emph{Experimental Psychology}, vol.~50, no.~1, pp. 19--25, 1955.

\bibitem{ellis:1998}
S.~R. Ellis and B.~M. Menges, ``Localization of virtual objects in the near
  visual field,'' \emph{Human Factors}, vol.~40, no.~3, pp. 415--431, 1998.

\bibitem{mccandless:2000}
J.~W. McCandless, S.~R. Ellis, and B.~D. Adelstein, ``Localization of a
  time-delayed, monocular virtual object superimposed on a real environment,''
  \emph{Presence: Teleoperators and Virtual Environments}, vol.~9, no.~1, pp.
  15--24, 2000.

\bibitem{singh:2010}
G.~Singh, J.~E. {Swan~II}, J.~A. Jones, and S.~R. Ellis, ``Depth judgment
  measures and occluding surfaces in near-field augmented reality,'' in
  \emph{ACM Symp. on Applied Perception in Graphics and Visualization
  (APGV)}.\hskip 1em plus 0.5em minus 0.4em\relax New York, NY, USA: ACM, 2010,
  pp. 149--156.

\bibitem{rosa:2016}
N.~Rosa, W.~H{\"u}rst, P.~J. Werkhoven, and R.~C. Veltkamp, ``{Visuotactile
  Integration for Depth Perception in Augmented Reality},'' in \emph{Proc. of
  ACM Intern. Conf. on Multimodal Interaction (ICMI)}.\hskip 1em plus 0.5em
  minus 0.4em\relax ACM, 2016, pp. 45--52.

\bibitem{rolland:1995}
J.~P. Rolland, W.~Gibson, and D.~Ariely, ``Towards quantifying depth and size
  perception in virtual environments,'' \emph{Presence: Teleoperators and
  Virtual Environments}, vol.~4, no.~3, pp. 24--49, 1995.

\bibitem{pedhazur:1982}
E.~J. Pedhazur, \emph{Multiple Regression in Behavioral Research},
  2nd~ed.\hskip 1em plus 0.5em minus 0.4em\relax Holt, Rinehart and Winston,
  1982.

\bibitem{cohen:2003}
J.~Cohen, P.~Cohen, S.~G. West, and L.~S. Aiken, \emph{Applied Multiple
  Regression/Correlation Analysis for the Behavioral Sciences}, 3rd~ed.\hskip
  1em plus 0.5em minus 0.4em\relax Lawrence Erlbaum Associates, 2003.

\bibitem{balram:2014}
N.~Balram, ``The next wave of {3-D}---light-field displays,'' \emph{Information
  Display}, vol.~30, no.~6, pp. 4,49, November/December 2014.

\bibitem{wu:2014}
W.~Wu, K.~Berkner, I.~To\v{s}i\'{c}, and N.~Balram, ``Personal near-to-eye
  light-field displays,'' \emph{Information Display}, vol.~30, no.~6, pp.
  16--22, November/December 2014.

\bibitem{duchowski:2014}
R.~I. Wang, B.~Pelfrey, A.~T. Duchowski, and D.~H. House, ``{Online 3D Gaze
  Localization on Stereoscopic Displays.}'' \emph{ACM Trans. on Applied
  Perception}, vol.~11, no.~1, pp. 1--21, 2014.

\end{thebibliography}

\begin{IEEEbiography}{Gurjot Singh}
Biography text here.
\end{IEEEbiography}

\begin{IEEEbiography}{Stephen R. Ellis}
Biography text here.
\end{IEEEbiography}

\begin{IEEEbiography}{J. Edward Swan II}
Biography text here.
\end{IEEEbiography}

\end{document}